\documentclass[9pt,twocolumn,romanappendices]{IEEEtran}
\usepackage[numbers,square,comma,sort&compress]{natbib}
\usepackage{graphicx,amsfonts,amsmath,amssymb,array,url}
\usepackage{subfigure,setspace,multirow}
\usepackage{epstopdf}

\usepackage[numbers,square,comma,sort&compress]{natbib}

\def\proof{\noindent\hspace{2em}{\itshape Proof: }}
\def\endproof{\hspace*{\fill}~$\blacksquare$\par\endtrivlist\unskip}
\newtheorem{theorem}{Theorem}
\newtheorem{lem}{Lemma}

\usepackage{color}
\usepackage{cite}
\usepackage{hyperref}
\begin{document}

\title{\huge On the Security Analysis of a Cooperative Incremental Relaying Protocol in the Presence of an Active Eavesdropper}
\author{\large{Saeed Vahidian, Sajad Hatamnia, and Benoit Champagne,~\IEEEmembership{Senior Member,~IEEE}}
\thanks{ S. Vahidian is with the Department of Electrical and Computer Engineering, University of California San Diego (UCSD), CA, USA (e-mail: saeed@ucsd.edu). S. Hatamnia is with the Faculty of Electrical and Computer Engineering, K. N. Toosi University of Technology, Tehran, Iran (e-mail: sajad.hatamnia@gmail.com). B. Champagne is with the Department of Electrical and Computer Engineering, McGill University, Montréal, QC H3A 0E9, Canada (e-mail: benoit. champagne@mcgill.ca). }}

\maketitle


\vspace{-2mm}

\begin{abstract}
Physical layer security offers an efficient means to decrease the risk of confidential information leakage through wiretap links. In this paper, we address the physical-layer security in a cooperative wireless subnetwork that includes a source-destination pair and multiple relays, exchanging information in the presence of a malevolent eavesdropper. Specifically, the eavesdropper is active in the network and transmits artificial noise (AN) with a multiple-antenna transmitter to confound both the relays and the destination. We first analyse the secrecy capacity of the direct source-to-destination transmission in terms of intercept probability (IP) and secrecy outage probability (SOP). A decode-and-forward incremental relaying (IR) protocol is then introduced to improve reliability and security of communications in the presence of the active eavesdropper. Within this context, and depending on the availability of channel state information, three different schemes (one optimal and two sub-optimal) are proposed to select a trusted relay to improve the achievable secrecy rate. 
For each one of these schemes, and for both selection and maximum ratio combining at the destination and eavesdropper, we derive new and exact closed-form expressions for the IP and SOP. Our analysis and simulation results demonstrate the superior performance of the proposed IR-based selection schemes for secure communication. They also confirm the existence of a floor phenomenon for the SOP in the absence of AN.
\end{abstract}


Artificial noise, incremental relaying, network, physical-layer security. 

\IEEEpeerreviewmaketitle
\markboth{ {IEEE T\MakeLowercase{ransactions on} I\MakeLowercase{nformation} F\MakeLowercase{orensics} \MakeLowercase{and} S\MakeLowercase{ecurity}}}{}

\def\baselinestretch{.9}
\vspace{-1mm}
\section{Introduction}
Wireless communication is naturally susceptible to eavesdropping due to the openness of the wireless medium and its broadcast nature. Therefore, confidential information exchanged between legitimate wireless nodes may easily be intercepted by unauthorized users. Due to increasing demand for private communication over wireless channels, security issues in wireless networks have gained considerable interest in recent years. Traditionally, security is implemented via cryptographic protocols using public or private keys at upper layers of the network stack. However, due to vulnerability in secret key distribution and management in dense wireless networks, information could be decrypted readily if the eavesdropper obtains the encryption key.
\subsection{Background}
Using an information-theoretic approach, Shannon~\citep{Shannon:1949} and Wyner~\citep{Wyner:1975}, and shortly afterwards Csisz{\'a}r and K{\"o}rner~\citep{Csiszar:Trans:1978}, have argued that it is possible to achieve perfectly secure communications without the use of cryptographic schemes if the channel of the wiretap link is inferior in quality to the legitimate channel. In that case, a confidential message can be encoded such that it can be reliably decoded at its intended destination while revealing almost no information to the eavesdropper. On this basis, physical (PHY) layer security derived from the information-theoretic perspective has attracted much attention recently as a promising approach for protecting against eavesdropping, without significantly increasing computational complexity~\citep{ElGamal:TIF:2012, gamal1, SaeedH2, SaeedH1, gamal2}. The basic idea is to exploit the PHY characteristics of the wireless channels in order to mitigate eavesdropping attacks. This line of work was extended in~\citep{Cheong:Thesis:1976}, where the impact of feedback on a wiretap channel was examined in terms of secrecy capacity, revealing that secure communication is still feasible, even when the wiretap link is superior to the legitimate channel by exploiting feedback information.

Taking advantage of multi-antenna systems to combat wireless fading as well as increasing link reliability and secrecy capacity, there has been a growing interest in extending the basic Gaussian wiretap channel to multiple-input multiple-output (MIMO) terminals~\citep{moh4, Soysal:TVT:2013, moh3}. The authors in~\citep{Khisti:TIT:2010} focused on the achievable secrecy capacity of a multi-input single-output (MISO) configuration, while in~\citep{Huang:TSP:2011} the PHY layer security in MIMO relay networks was studied, revealing a significant improvement in terms of secrecy rate through the use of MIMO relays. The secrecy capacity of a broadcast MIMO wiretap channel for an arbitrary number of transmit/receive antennas was studied in~\citep{Hassibi:TIT:2011}, which showed that the perfect secrecy capacity is equal to the difference in mutual information between the wiretap and legitimate links.
 However, considering the hardware cost and size limitations of multiple-antenna systems, cooperative relaying offers a compelling alternative that enables single-antenna nodes to enjoy the benefits of multiple-antenna systems while enhancing end-to-end security and reliability of communications~\citep{Vahidian:WCL:2015,Petropulu:TSP:2010, Shen:Conf:2013, moh2}.

Depending on the role played by the relay in cooperative schemes, three different generic scenarios can be identified. In the first scenario, the relay nodes aim to assist the eavesdropper by decreasing the secrecy rate~\citep{Erkip:Trans:2011}. In the second scenario, the relay acts as both a collaborator and an eavesdropper~\citep{Oohama:2006}. In the third scenario, which is the focus of this work,  the relay collaborates with the source to enhance security of the legitimate link~\citep{Fawaz:TWC:2015}.
Most of the existing works on user cooperation for PHY layer security focus on developing the secrecy rate from an information-theoretic viewpoint. Three different types of cooperative schemes are investigated, namely: amplify-and-forward (AF)~\citep{moh1}, decode-and-forward (DF)~\citep{Petropulu:TSP:2010} and cooperative-jamming (CJ)~\citep{Petropulu:TSP:2010} are the most important protocols. In particular, optimal relay weight selection and power allocation strategies are proposed to enhance the achievable secrecy rate for the second hop. The authors in~\citep{ ElGamal:TIF:2008} study the four-node (i.e., source, destination, relay and eavesdropper) secure communication system for different relay strategies, including DF and noise-forwarding (NF). In~\citep{Jeong:TSP:2011}, the four-node system is further examined in the context of multi-carrier transmissions, where the aim is to maximize the sum secrecy rate under a total system power constraint. A novel relay selection strategy with jamming is investigated in~\citep{Thompson:TWC:2009}, where the aim is to improve security at the destination under the assumption that the eavesdropper only overhears the second hop.

Reference~\citep{Liang:SPL:2015} analyzes secure relay and jammer selection for the PHY-layer security improvement of a wireless network including multiple intermediate nodes and eavesdroppers. In~\citep{ElGamal:TIF:2012}, the authors propose a new multi-hop strategy where the relays add independent randomization in each hop, which leads to significant secrecy improvement for the end-to-end transmission. The PHY layer security is further explored in~\citep{Chen:TIF:2012} for the two-way relay channels, where multiple two-way relays are employed to enhance the secrecy rate against eavesdropping attacks. Other related works addressing the problem of PHY layer security in the presence of multiple intermediate nodes or eavesdroppers include~\citep{ IESe, qaa, qaaa, qaaaa, qbb, qbbb, qbbbb}.


   The aforementioned works are limited to cases where the eavesdropper node can only overhear the source's message or that of the relay but not both. The sub-network models invoked in these and other studies are often afflicted by further restrictions, which may limit their realm of application in practice. This includes the following: consideration of a single eavesdropper equipped with single antenna, as opposed to multiple antennas; legitimate network sending artificial noise to degrade the wiretap link but not the converse; and adoption of conventional cooperation protocols which are not spectrally efficient.

\subsection{Technical Contributions}
 Motivated by these observations, in this paper, we investigate the effects of different relay selection schemes as well as different combining techniques (under the Rayleigh fading model) on the PHY layer security when the eavesdropper has access to both the source and relay messages. Three relay selection schemes are employed based on the availability of the channel state information (CSI), namely: conventional selection, minimum selection, and optimal selection. In conventional selection, the selected relay is the one that results in the highest SNR at the destination.  In minimum selection, the selected relay is the one that results in the lowest SNR at the eavesdropper.  Finally, in optimal selection, the selected relay is the one that maximizes the secrecy capacity.

For each one of these schemes, we study the performance of a DF-incremental relaying (IR) protocol in the presence of eavesdropper generated AN at both the relays and the destination. Cooperative schemes based on IR outperforms those based on the traditional retransmission of the source message~\citep{ Larsson:TWC:2009}. In effect, they are amongst the best performing schemes, as they preserve the channel resources \emph{i.e.}, bandwidth and energy, while maintaining reliable communication.

   By employing IR and due to the presence of direct links, the destination and the eavesdropper each receive two different versions of the source message. Consequently, diversity signal combining techniques can be employed by these nodes, including: selection combining (SC), which only selects the best signal out of all replicas for further processing; and maximal ratio combining (MRC), which coherently adds the signal replicas together for detection. For convenience, henceforth, we shall use the nomenclature in Table~1 to refer to the various combinations of relay selection and signal combining schemes. In this table each scheme is identified by a three-letter label where the first letter stands for the DF strategy, the second letter represents the type of the signal combining technique and the third letter denotes the adopted relay selection scheme. In addition, the scheme labeled ``DT" denotes the conventional direct transmission and finally, ``All relays" means that all successful relays in decoding cooperate simultaneously in the next phase without employing any relay selection scheme.

\begin{table}
\begin{center}
\caption{Adopted Nomenclature for Relay Selection Schemes under Study}
\vspace{-1mm}
\begin{tabular}{| c| c| c| c| c| c| c| c| c| c}
\hline
Scheme & Signal Combining  & Relay Selection \\
\hline
  DMC & MRC  & Conventional \\
  \hline
  DSC & SC & Conventional  \\
 \hline
  DMM & MRC  & Minimum \\
  \hline
  DSM  & SC & Minimum  \\
  \hline
  DMO & MRC &  Optimal  \\
  \hline
  DSO & SC &  Optimal  \\
  \hline
  DMA & MRC & All relays \\
  \hline
  DSA & SC & All relays   \\
  \hline
  DT & - & No relay \\
  \hline
\end{tabular}
\vskip -15pt
\label{table:nonlin}
\end{center}
\end{table}

While the literature on PHY layer security is abundant, the study of security issue for cooperative IR networks affected by eavesdropper generated AN has not been previously addressed. Specifically, our work differs from the aforementioned studies in many aspects, its main contributions being summarized as follows:
\begin{itemize}
\item  We consider a cooperative wireless network with multiple relays in the presence of an active eavesdropper and investigate  communication security from the perspective of information theory. Unlike previous works, i.e.~\citep{Zurita:SPL:2012, McKay:TVT:2013, 6951465, 7752562, Wang17, 7417842} where the source or relays transmit AN together with information signals to deliberately interfere with the eavesdropper's received signal, both the relays and the destination node in our model are confounded by AN originating from the eavesdropper node, which represents the worst case scenario.
\item Cooperative diversity with traditional fixed relaying leads to a notable loss in the system capacity and efficiency because it requires two time intervals for half-duplex transmission. In order to prevent such a loss, we consider a novel IR strategy and investigate its performance in the context of secure communications.
\item We present and investigate three different relay selection schemes to enhance PHY layer security against eavesdropping attack. In contrast to~\citep{Shen:TVT:2014} which assumes conventional relay selection and therefore only considers the relay-destination links, we herein depending on the availability of CSI examine alternative selection schemes which take into account the quality of both source-relay and relay-destination channels.
\item  We derive closed-form expressions for the intercept probability (IP) and the secrecy outage probability (SOP) of all the proposed schemes for cooperative IR networks, thereby fully characterizing the associated security-reliability trade-off~\citep{B.Champagne:TCOM:2015}. To provide further insight into system behavior, we also derive corresponding asymptotic expressions for the SOP of each scheme in the high SNR regime. These expressions facilitate system design and help better understand the role played by various internal parameters and their interaction.
\end{itemize}

\subsection{Key Findings}

{Based on the analysis provided in this paper for the cooperative IR wireless network in the presence of an active eavesdropper, our key results include:}
\begin{itemize}
\item  When perfect CSI is available, the optimal relay selection scheme provides the best security performance as compared to the others. This follows because the optimal scheme takes into account the quality of both the source-destination and relay-destination links in its decision metric. In addition, the conventional selection scheme always outperforms minimum selection, which can be justified by invoking the concept of diversity order. Indeed conventional selection provides a diversity gain for the legitimate links when compared to minimum selection.
\item The performance of SC is worse than that of MRC, typically exhibiting a few dBs of power penalty. This is the price paid for reduced complexity with SC, which allows a trade-off between complexity and performance. SC has its own merits, which explains its wide application  in wireless communication in general, as well as in works dealing with security (see \emph{e.g.},~\citep{Shen:TVT:2014}). However, there is not a significant performance gap between the two combining techniques when the minimum selection scheme is employed.
\item Only marginal performance improvements can be obtained by increasing the number of relays for the DMM and DSM schemes.
\item  In the high SNR regime, all the proposed schemes achieve the same diversity gain, while the difference in their performance can be characterized by their achieved coding gain.
 \end{itemize}

\textit{Mathematical Notations:}
The notation $o\left( x \right)$ means an higher order term in $x $, (i.e. $ \mathop {{\rm{lim}}}\limits_{x \to 0} o(x) / x =0$); $f(x)$ and $F(x)$ respectively denote the probability distribution function (PDF) and cumulative distribution function (CDF) of random variable (RV) $X$; $\Gamma \left( {a,x} \right)$ is the upper incomplete gamma function while $\Phi \left( {a,b;x} \right)$ is the confluent hypergeometric function of the second kind.

The rest of the paper is organized as follows. The adopted system and channel models for cooperative relaying with active eavesdropper are discussed in Section II. Sections III and IV present the proposed IR-based schemes and their secrecy performance analysis, respectively. Section V analyzes diversity order in the high SNR regime. Section VI presents selected numerical simulation results to support the theoretical study. Finally, Section VII contains concluding remarks.

\section{System Model}

Consider the generic topology shown in Fig. 1 for secure communication in a cooperative wireless sub-network consisting of a source $S$, a destination $D$, and a cluster of $M$ DF relays $R_m, m\in\left\{ {1,...,M} \right\}$. The purpose of the relays is to assist the data transmission between the source and the destination,  in order to protect against the overhearing attack of a malicious eavesdropper $E$. The source, destination and relay nodes are characterized by the half-duplex constraint and therefore cannot transmit and receive simultaneously, while the eavesdropper node works in full duplex mode. Specifically, the eavesdropper is active and utilizes a hybrid overhearing and AN generating mechanism. Herein, ``hybrid" means that during the data exchange, the eavesdropper not only overhears to extract confidential information but also propagates AN to degrade the PHY layer security of the legitimate sub-network. To this end, the eavesdropper can use multiple-antennas or collude with other attackers concealed nearby to generate AN and confound the target receivers\footnote{Even if the eavesdropper is equipped with a single antenna, by collaborating with helper nodes in its surrounding, it can control the generation of AN that can still be malicious. For instance, multiple eavesdroppers can collude to form an antenna array.}.

\begin{figure}
\centering
\includegraphics[width=3.5 in]{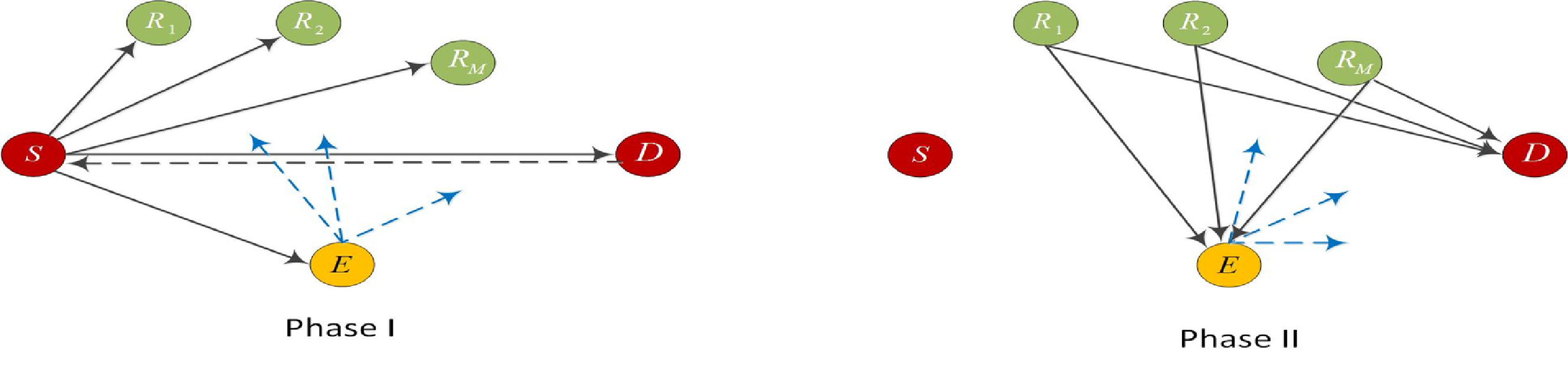}
\vspace{-1em}
\caption{A wireless relay network consisting of one source ($S$), one destination ($D$), and $M$ relay nodes $R_m$, exchanging information in the presence of an eavesdropper ($E$). The continuous and dashed black lines correspond to the legitimate forward and backward links of the legitimate network, respectively, while the dashed blue line illustrates the AN propagated by $E$.} \vspace{-4mm}
\label{fig:AchievableRate_secondScenario}
\end{figure}


Unlike traditional cooperation~\citep{Bao:TWC:2013, 7996473}, in the considered topology only a subset of the $M$ relays will be activated. Specifically, we consider IR as a cooperation protocol which exploits an one-bit feedback from the destination to the source in the form of Acknowledgement/Negative-Acknowledgement (ACK/NACK) signaling as shown in Fig.~1. In the proposed model, in the first phase, $S$ broadcasts its signal and all the relays $R_m$, $m \in \{1,...,M\}$, attempt to decode it. Let $\cal{F}$ denote the random subset of relays that can successfully decode the source message, referred to as the well-informed relay subset (WIRS). Accordingly, the sample space of all the possible WIRS outcomes is the power set ${\cal{P}}(\{R_1,...,R_M\})$ with cardinality $2^M$. In the sequel, it is convenient to individually represent these subsets by ${\cal{F}}_n$ where the index $n \in \{1,2, 3,...,2^M\}$. Next, let ${\cal{R}}$ be a pre-determined rate which is contingent on the quality of service (QoS) of the source-destination link. On the basis of the rate ${\cal{R}}$, the destination decides whether another copy of the data signal is required or not. As previously mentioned, the retransmission process is based on an ACK/NACK mechanism, in which short-length error-free packets are broadcasted by the destination $D$ over a separate narrow-band channel, in order to inform the source and the relay nodes of that reception QoS status. In the second time slot, if necessary, i.e., if the rate of the source-destination channel falls below $\cal{R}$, the best relay processes the received signal using the DF protocol~\citep{hatamnia2017network, Vahidian:2015:WPC}, whereby a copy of the original source message is generated and transmitted again to the destination.

%


It is assumed that all wireless links in Fig.~1, including the $E$'s channels, exhibit frequency flat Rayleigh block fading. This means that the fading channel coefficients remain (approximately) static for one coherence interval, but change independently in different coherence intervals according to a zero-mean circularly symmetric complex Gaussian distribution. We let $h_{i,j}$ denote the complex valued channel coefficient characterizing the transmission from node $i$ to node $j$,  where $i, j \in \left\{ {s, d, e, r_1,...,r_M} \right\}$. The receivers at nodes $S$, $D$, $E$, and $R_m$ are impaired by additive white Gaussian noise (AWGN). Hence, the signal-to-noise ratio (SNR) with respect to (w.r.t.) link $i$-$j$ follows an exponential distribution with mean denoted as ${\bar \sigma _{ij}}$.

\subsection{Direct transmission}
In the following and for later reference, we proceed by presenting the security analysis in the special case of direct transmission (i.e., without using relay cooperation) as a benchmark. Subsequently, in Section III, we will propose the DF-based IR protocol to improve the PHY layer security against eavesdropping attack, and extend our analysis to this latter case.


\begin{figure}
\centering
\includegraphics[width=2.5 in]{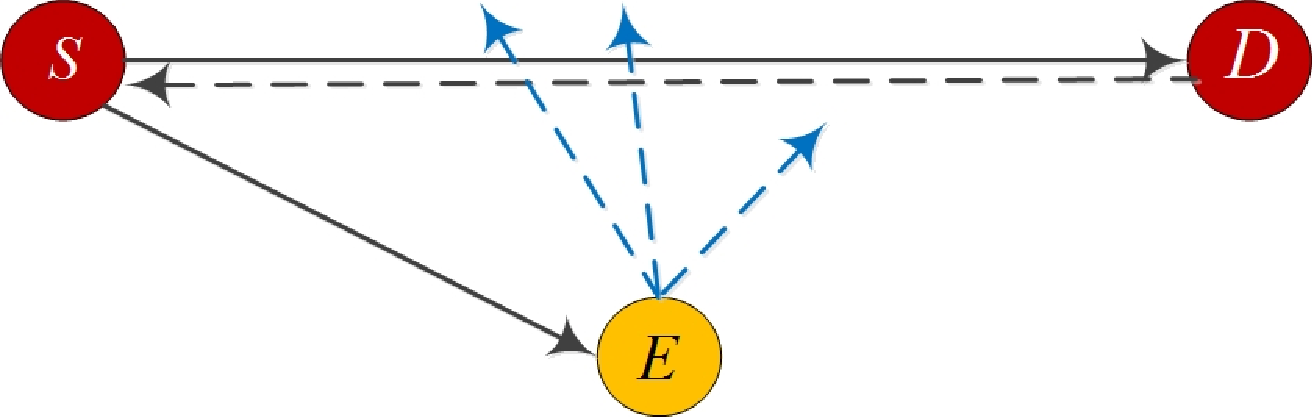}
\vspace{-1em}
\caption{A wireless relay network consisting of one source ($S$), one destination ($D$),  and exchanging information in the presence of an eavesdropper ($E$). The continuous and dashed black lines correspond to the forward and backward links of the legitimate network, while the dashed blue line shows the AN propagated by $E$.} \vspace{-4mm}
\label{fig:AchievableRate_secondScenario}
\end{figure}


In the direct transmission case (Fig. 2), source $S$ transmits a sequence of complex valued digital symbols to destination $D$, at the rate $\cal{R}$ in units of bits per channel use. Here, we assume that quadrature phase-shift keying (QPSK) is employed as the modulation technique, and we let $\cal{A}$ with cardinality $Q=|{\cal{A}}|$ denote the normalized symbol constellation. At a given time instant (i.e., channel use), $S$ transmits a scaled random symbol $\sqrt{P_s}s $, where $s \in \cal{A}$ with $E\{|s|^2\}=1$ and $P_s$ is the source transmit power. Due the broadcast nature of electromagnetic waves, the radio signal transmitted by $S$ to $D$ will also reach some unintended areas, resulting in information leakage. Consequently, eavesdropper $E$ may overhear the transmission of $S$ and reproduce the original confidential signal.

Meanwhile, the AN vector expressed by $[\sqrt{P_1}x_1,\ldots,\sqrt{P_N}x_N]$ is emitted by node $E$, where $N$ is the number of available transmit antennas, $x_i$ for $i\in\{1,2,..., N\} $ are random variables taken from a complex circular Gaussian distribution with zero mean and variance $E\{|x_i|^2\}=1$, and $P_i$ is the corresponding transmit power allocated to the $i^{\rm{th}}$ antenna. Ideally, the AN is generated to be in the null space of node $E$'s receiving channel, and thus, does not affect $E$ but degrades the receivers' channels~\citep{Negi:TWC:2008}. Hence, the received signals at $D$ and $E$ are respectively given by
\begin{align}
&r_{s,d} = \sqrt {{P_s}} {h_{s,d}}s + \sum\limits_{i = 1}^{{N}} {\sqrt {{P_{i}}} {c_{i,d}}} {x_{i}}\, + {n_{s,d}}, \\&
{r_{s,e}} = \sqrt {{P_s}} {h_{s,e}}s + {n_{s,e}},
\end{align}
\noindent where $c_{i,d}$ is the complex channel gain between the $i^{\rm{th}}$ antenna of $E$ and the destination $D$, while $n_{s,d} \sim \mathcal{CN}\left( {0,{\sigma _{n}^2}} \right)$ and ${n_{s,e}} \sim \mathcal{CN}\left( {0,{\sigma _{n}^2}} \right)$ are the additive noise terms at the destination and the eavesdropper terminal respectively~\footnote{For simplicity, we assume that these noise terms have the same power but the analysis can be extended to the case of different noise powers.}. Then, the channel capacity of the direct $S\hspace{-1.1mm}-\hspace{-1.1mm}D$ link and the achievable rate of the wiretap $S\hspace{-1.1mm}-\hspace{-1.1mm}E$ link are given by
\begin{align}
&{{C}}_{sd}^{{\rm{}}} = {\log _2}\left( {1 + {\Psi _{sd}}} \right),~~~~~~~{\Psi _{sd}} = \frac{{{P_{\rm{s}}}{{\left| {{h_{s,d}}} \right|}^2}}}{{\sum\limits_{i = 1}^{{N}} {{P_{i}}{{\left| {{c_{i,d}}} \right|}^2}}  + \sigma _n^2}},\\&
{C_{se}^{\rm{}}} = {\log _2}\left( {1 + {\Psi _{se}}} \right),~~~~~~~~~~~{\Psi _{se}} = \frac{{{P_{\rm{s}}}{{\left| {{h_{s,e}}} \right|}^2}}}{{\sigma _n^2}}.
\end{align}

According to~\citep{Shannon19488}, when the capacity of the wiretap $S\hspace{-1.1mm}-\hspace{-1.1mm}E$ link is lower than the data rate $\cal{R}$, $E$ will fail to decode the message from $S$, while the legitimate $S\hspace{-1.1mm}-\hspace{-1.1mm}D$ link remains secure. However, if the capacity of the wiretap link becomes higher than the data rate $\cal{R}$, $ E$ may succeed in decoding $S$'s message and hence, an intercept event occurs. Within this context, the intercept probability (IP) defined below is a key metric in evaluating the performance of PHY layer security: 
\begin{equation}
{\cal{P}}_{{\rm{int}}}^{{\rm{DT}}}=\Pr \left( {C_{se}^{{\rm{}}} > {\cal{R}}} \right)=\exp \left( { - \frac{\delta }{{{{\bar \sigma }_{se}}}}} \right)
\label{C_SD_sec_DF_DT}
\end{equation}
\noindent where $\delta =2^{{\cal{R}}}-1$ and the superscript DT stands for direct transmission. As expected, the IP is contingent on the transmit power of source $S$ and the quality of the wiretap $S\hspace{-1.1mm}-\hspace{-1.1mm}E$ link, through the parameter $\bar{\sigma}_{se}$, as well as the data rate ${\cal{R}}$. Note that increasing the data rate or decreasing the transmit power of the source, causes the IP to decrease, which in turn improves the security of the network. However, this comes at the cost of a deterioration in transmission reliability, since the SOP of the legitimate link increases (see below) when a higher data rate or lower transmit power is adopted at $S$.

Let us next investigate the achievable secrecy rate of direct transmission, which is defined as the difference between the information rate of the $S\hspace{-1.1mm}-\hspace{-1.1mm}D$ link and that of the $S\hspace{-1.1mm}-\hspace{-1.1mm}E$ link:
\begin{equation}
C_{sd}^{{\rm{DT}}}=\left[ {C_{sd}^{{\rm{}}} - C_{se}^{{\rm{}}}} \right]^+ = {\log _2}\left( {\frac{{1 + {\Psi _{sd}}}}{{1 + {\Psi _{se}}}}} \right),
\label{C_SD_sec_DF_DT}
\end{equation}



\noindent where ${[x]^+} = \max \left[ {x,0} \right]$~\footnote{When the secrecy capacity is negative, a SOP event occurs.}. Under the security constraint, the legitimate network is in outage whenever a transmitted message cannot be received reliably. Specifically, for a given secure rate $\cal{R}$, a secrecy outage  event occurs when the secrecy rate falls below the thresholding $\cal{R}$. In this regard, the secrecy outage probability (SOP) provides another key metric in evaluating the performance of PHY layer security.
For the traditional direct transmission mode, the SOP can be formulated as
\begin{align}
{\cal{P}}_{{\rm{out}}}^{{\rm{DT}}} =&
\Pr \left( {C_{{\rm{sd}}}^{{\rm{DT}}} \le {\cal{R}}} \right) = \Pr \left( {\frac{}{}{{\log }_2}\left( {\frac{{1 + {\Psi _{sd}}}}{{1 + {\Psi _{se}}}}} \right) < {\cal{R}}} \right)\nonumber\\
=& \Pr \left( {{\Psi _{sd}} \le {2^{{\cal{R}}}}{\Psi _{se}} +\delta} \right).
\end{align}
The CDF of RV ${{\Psi _{sd}}}$ and pdf of RV ${{\Psi _{se}}}$ at hand can be expressed as \citep{Fawaz:TWC:2015}
\begin{align}
&{F_{{\Psi _{sd}}}}\left( {{\Psi _{sd}}} \right) = 1 - \sum\limits_{i = 1}^N {{{\rm{\pi }}_{sd}}} \frac{{{\kappa _{sd}}}}{{{\Psi _{sd}} + {\kappa _{sd}}}}\exp \left( { - \frac{{{\Psi _{sd}}}}{{{{\bar \sigma }_{sd}}}}} \right), \\&
{f_{{\Psi _{se}}}}\left( {{\Psi _{se}}} \right) = \frac{1}{{{{\bar \sigma }_{se}}}}\exp \left( { - \frac{1}{{{{\bar \sigma }_{se}}}}{\Psi _{se}}} \right),
\end{align}
\noindent where ${\kappa _{sd}} = \frac{{{{\bar \sigma }_{sd}}}}{{{{\bar \sigma }_{id}}}}$\footnote{$ {\left| {{c_{i,j}}} \right|^2}, j \in \left\{ {m,d} \right\}$ follows an exponential distribution with mean ${\bar \sigma }_{ij}$.} and ${{\rm{\pi }}_{sd}} = \prod\limits_{\mathop {i = 1}\limits_{j \ne i} }^{{N}} {\frac{{{{\bar \sigma }_{id}}}}{{{{\bar \sigma }_{id}} - {{\bar \sigma }_{jd}}}}}$. Using these expressions, the SOP of the conventional direct transmission is obtained as
\begin{align}\label{Pr_SD_DF}
{\cal{P}}_{{\rm{out}}}^{{\rm{DT}}} &= \int_0^\infty  {{F_{{\Psi _{sd}}}}\left( {{{{2^{\cal{R}}}}}{\Psi _{se}} + \delta } \right)} {f_{{\Psi _{se}}}}\left( {{\Psi _{se}}} \right)\,\,d{\Psi _{se}} \notag \\
&
=
 1 - \sum\limits_{i = 1}^N \bar \sigma _{se}^{ - 1}{{\pi _{sd}}} {\kappa _{sd}}{\left( {\delta  + 1} \right)^{ - 1}}\exp \left( {\frac{{\left( {\delta  + {\kappa _{sd}}} \right)}}{{{{\bar \sigma }_{se}}\left( {\delta  + 1} \right)}} + \frac{{{\kappa _{sd}}}}{{{{\bar \sigma }_{sd}}}}} \right)\nonumber\\
&\times\Gamma \left( {0,\frac{{\left( {\delta  + {\kappa _{sd}}} \right)}}{{{{\bar \sigma }_{se}}\left( {\delta  + 1} \right)}} + \frac{{\left( {\delta  + {\kappa _{sd}}} \right)}}{{{{\bar \sigma }_{sd}}}}} \right).
\end{align}


%
\section{Proposed DF Incremental Relaying Scheme}
In this section, the proposed relay selection scheme is exposed and its secrecy performance analyzed.  We here make use of the DF scheme along with an IR protocol to augment the spectral efficiency over fixed relaying systems. Based on the model description provided in Section II, the received signal at $R_m$ and subsequently received signal at $D$ and $E$ from the selected relay, are respectively given by
\begin{align}
&{r_{s,m}} = \sqrt {{P_s}} {h_{s,m}}s + \sum\limits_{\ell  = 1}^{{N}} {\sqrt {{P_{\ell }}} {c_{\ell,m }}} {x_{\ell }} + {n_{s,m}},\label{r_k_DD}
\\&
{r_{m,d}} = \sqrt {{P_m}} {h_{m,d}}s + \sum\limits_{i = 1}^{{N}} {\sqrt {{P_{i}}} {c_{i,d}}} {x_{i}} + {n_{m,d}},\label{r_k_D}
\end{align}
\vspace{-2.5mm}
\begin{equation}
{r_{m,e}} = \sqrt {{P_m}} {h_{m,e}}s + {n_{m,e}},
\end{equation}

\noindent where $P_s=P'/2$ and $P_m=P'/2$ are the transmitted power at $S$ and $ R$ respectively and $P'$ is the total power budget of the network. In addition, ${c_{\ell,m}}$ is the complex channel gain between the $\ell$th antenna of $E$ and the $m$th relay while $n_{s,m} \sim \mathcal{CN}\left( {0,{\sigma _{n}^2}} \right)$, $n_{m,d} \sim \mathcal{CN}\left( {0,{\sigma _{n}^2}} \right)$ and $n_{m,e} \sim \mathcal{CN}\left( {0,{\sigma _{n}^2}} \right)$ are additive noise terms. There are two possible cases for the data transmission depending on whether the WIRS $\mathcal{F}$ is empty or not. For simplicity, let $\mathcal{F}= \emptyset$ represents the former case and $\mathcal{F}= \mathcal{F}_n$ the latter.
\begin{itemize}
\item Case $\mathcal{F}= \emptyset$: This case corresponds to a situation where all the candidate relays fail in perfectly decoding the source signal. From an information theoretic perspective,  this condition can be expressed based on~\eqref{r_k_DD} in the following form,
\begin{equation}
\frac{1}{2}{\log _2}\left( {1 + {\Psi _{sm}}} \right) < {\cal{R}},~~~\forall m,  \label{D=0}
\end{equation}
\noindent where ${\Psi _{sm}} = \frac{{{P_{\rm{s}}}{{\left| {{h_{s,m}}} \right|}^2}}}{{\sum\limits_{i = 1}^N {{P_i}{{\left| {{c_{i,m}}} \right|}^2}}  + \sigma _n^2}}$. Based on \eqref{D=0}, the occurrence probability of case $\mathcal{F}= \emptyset$ is given by
\begin{align}
\Pr \left( \mathcal{F= \emptyset}\right) &=\!\!\prod\limits_{m = 1}^M\!\! {\Pr \left( {\frac{1}{2}{{\log }_2}\left( {1 + {\Psi _{sm}}} \right) < {\cal{R}}} \right)}\nonumber\\
&=
\!\!\prod\limits_{m = 1}^M \!\!{\Pr \left( {{\Psi _{sm}} < {2^{2{\cal{R}}}} - 1} \right)  }
\!=\!\!\prod\limits_{m = 1}^M { {\cal{E}}_{\ell m} },\!
\end{align}
\noindent where ${\cal{E}}_{\ell m}=1 - \sum\limits_{\ell  = 1}^{{N}} {{{\rm{\pi }}_{s }}} \frac{{{\kappa _{s }}}}{{\varrho  + {\kappa _{s }}}}\exp \left( { - \frac{\varrho }{{{{\bar \sigma }_{sm}}}}} \right)$, $\varrho= 2^{2\cal{R}}-1$, ${{\rm{\pi }}_{s }} = \prod\limits_{\mathop {\ell = 1}\limits_{q \ne \ell } }^{{N}} {\frac{{{{\bar \sigma }_{\ell m }}}}{{{{\bar \sigma }_{\ell m }} - {{\bar \sigma }_{mq}}}}}$ and ${\kappa _{s}} = \frac{{{{\bar \sigma }_{sm}}}}{{{{\bar \sigma }_{\ell m }}}}$.

\item  Case $\mathcal{F}= \mathcal{F}_n$: This case corresponds to all the relays in the WIRS $\mathcal{F}_n$ being able to decode the source signal successfully, i.e., the event $\mathcal{F}= \mathcal{F}_n$ is formulated as
\begin{align}
&\frac{1}{2}{\log _2}\left( {1 + {\Psi _{sm}}} \right)> {\cal{R}} \rm{~if~and~only~if}~\it{m} \in{\cal{ F}}_n \label{D=D_m_1}
\end{align}

%
From \eqref{D=D_m_1}, the occurrence probability of case $\mathcal{F}= \mathcal{F}_n$ can be formulated as
\begin{align}
\Pr \left( {\mathcal{F} = {\mathcal{F}_n}} \right) =&
\prod\limits_{m \in {{\cal{F}}_n}} {\Pr \left( {\frac{1}{2}{{\log }_2}\left( {1 + {\Psi _{sm}}} \right) > {\cal{R}}} \right)}\times\nonumber\\
&\prod\limits_{m \in {{{\bar{\cal F}}_n}}} {\Pr \left( {\frac{1}{2}{{\log }_2}\left( {1 + {\Psi _{sm}}} \right) < {\cal{R}}} \right)}\notag\\
 =& \prod\limits_{m \in {{\cal{F}}_n}} {\Pr \left( {{\Psi _{sm}} > \varrho } \right)}
   \prod\limits_{m \in {{\bar{\cal F}}_n}} {\Pr \left( {{\Psi _{sm}} < \varrho } \right)}\nonumber\\
=& \prod\limits_{m \in {{\cal{F}}_n}} { {\cal{E}}_{\ell m}}   \prod\limits_{m \in {{\bar{{{\cal{F}}_n}}}}} \left({1- {\cal{E}}_{\ell m}} \right).
\end{align}
\end{itemize}

During the cooperative phase and according to the assumed security protocol, when ${\cal{F}}_n$ is non-empty, the best relay is chosen from ${\cal{F}}_n$ to forward its decoded signal toward the destination, allowing the eavesdropper to intercept the transmission.

\section{Security-Reliability Analysis of IR Schemes}

Based on the available knowledge of CSI for the different links and system complexity, different relay selection schemes are presented and analyzed for the following three cases. For Case I, corresponding to the situation when the CSI of the legitimate channels (i.e. $S-R_m, R_m - D$) is available but not that of the $R_m-E$ channel, the conventional  relay selection scheme is implemented. The latter scheme does only take into account the capacity of the legitimate channels, without considering the secrecy rate. In Case II, i.e. when the CSI of the $R_m-E$ link is known, minimum relay selection scheme is applied. We note that for Cases I and II, suboptimal relay selection is performed. Finally, in Case III, while the same CSI assumption as in Case II are made, an optimal relay selection scheme is implemented, whereby the relay that achieves the maximum secrecy rate is chosen for retransmission. Compared to conventional relay selection approaches~\citep{Bletsas2006, Zou.Trans.2010, saeedtvt2017, Bletsas:TWC:2007,} where only the CSI of the legitimate $S-R_m$ and $R_m-D$ links are required, in the optimal IR scheme, knowledge of the channel gain magnitudes of eavesdroppers channels is also needed for maximizing the secrecy rate through the use of an error-free feedback channel between $E$ and $D$~\citep{Ikhlef:GLOBE:2011, Shen:Conf:2013,Bloch:TIF:2008,Vahidian:WCL:2015}). The assumption of known eavesdropper CSI can be justified when the eavesdropper is seen as another legitimate user waiting to be served by the source while the latter is serving  the destination securely. Thus, the eavesdropper has to report its CSI to the source to be considered for future service. This is applicable particularly in networks combining multicast and unicast transmissions.

Without loss of generality, assuming that event $\mathcal{F}=\mathcal{F}_n$ occurs and relay ``$m$" is selected as the ``best one", the corresponding $R_m-D$ and $R_m-E$ channel capacities are
\vspace{-2mm}
\begin{align}
&C_{md}^{{\rm{DF}}} =  \frac{1}{2}{\log _2}\left( {1 + {\Psi _{md}}} \right), \label{eq:C-md}\\&
C_{me}^{{\rm{DF}}} = \frac{1}{2}{\log _2}\left( {1 + {\Psi _{me}}} \right),\label{C_k_e}
\end{align}
\noindent where ${\Psi _{md}} = \frac{{{P_m}{{\left| {{h_{m,d}}} \right|}^2}}}{{\sum\limits_{i = 1}^N {{P_i}{{\left| {{c_{i,d}}} \right|}^2}}  + \sigma _n^2}}$, ${\Psi _{me}} = \frac{{{P_m}{{\left| {{h_{m,e}}} \right|}^2}}}{{\sigma _n^2}}$ and the superscript DF refers to decode-and-forward. For the case of unsuccessful direct transmission, two time slots are required to transmit the data, justifying the factor of 1/2 in \eqref{eq:C-md}-\eqref{C_k_e}. To boost the effective channel gain and thereby enhance communication reliability, the destination then employs either MRC or SC of all signals received in both phases, and generates an estimation of the original signal after maximum likelihood decoding (MLD). As such, two categories of combining solutions do exist for both destination and eavesdropper nodes, leading to various cases in our analysis as detailed below (see also Table~1).




\subsection{Suboptimal Selection Case I: DMC}

Let us investigate the security and reliability performance of the DF IR scheme where a combination of the conventional relay mode and the MRC technique are applied. In this approach, the relay selection scheme does not take into account the eavesdropper's channels and the relay node is selected based on the instantaneous quality of the combined $S-R_m$ and $R_m-D$ link, with the aim to maximize the achievable rate at the destination node. Specifically, the index of the relay chosen to forward the legitimate signal from $D$ to $E$ is given by
\begin{equation}
{m^* }=\mathop {{\rm{arg}}{\rm{max}}}\limits_{m{\kern 1pt}  \in {\kern 1pt} {{\cal{F}}_n}} {\mkern 1mu}\; {C_{sd}^{\rm{DM}}},
\end{equation}
\noindent where we define $C_{sd}^{\rm{DM}}\hspace{-1mm} =\hspace{-1mm}  \frac{1}{2}{\log _2}\left( {1 + {\Psi _{md}} + {\Psi _{sd}}} \right) $. Notice that for the MRC technique, the instantaneous SNR is given by the sum of two SNRs, i.e.,  ${\Psi _{md}}$ and ${\Psi _{sd}}$. Then, the secrecy rate is expressed as
\begin{equation}
\!\!C_{sd{\rm{}}}^{{\rm{DMC}}}\hspace{-1mm}=\hspace{-1mm} {\left[ {{C_{d}^{\rm{DMC}}} \hspace{-1mm}-\hspace{-1mm} C_e^{{\rm{DM}}}} \right]^ + }\!\! \hspace{-1mm}=\hspace{-1mm} \frac{1}{2}{\log _2}\!\!\left(\! {\frac{{1 \hspace{-1mm}+\hspace{-1mm}  {\mathop {\max }\limits_{m\, \in \,{{\cal{F}}_n}}\!\! {\Psi _{md}}\hspace{-1mm}+\hspace{-1mm}{\Psi _{sd}}}}}{{1\hspace{-1mm} +\hspace{-1mm}   {{\Psi _{{m^* }e}}\hspace{-1mm}+\hspace{-1mm}{\Psi _{se}}} }}}\! \right)\!\!,
\end{equation}
\noindent where ${C_{d}^{\rm{DMC}}}\hspace{-1mm} = \mathop {{\rm{}}{\rm{max}}}\limits_{m{\kern 1pt} \in {\kern 1pt} {{\cal{F}}_n}} {C_{sd}^{\rm{DM}}}$  and  $C_e^{{\rm{DM}}}=\hspace{-1mm} \frac{1}{2}{\log _2}\left( {1 + {\Psi _{me}} + {\Psi _{se}}} \right)$.

We now derive an analytical expression for the IP for conventional relaying with MRC. To this end, we first present the following general expression for the IP, based on the law of total probability, that is applicable to various combinations of signal combining and relay selection schemes,
\begin{equation}
{{\cal{P}}_{{\rm{int}}}} = \Pr \left( {{\cal F} = \emptyset } \right){\cal{P}}_{{\rm{int}}}^{{\rm{DT}}}+ \sum\limits_{n = 1}^{{2^M} - 1} {\Pr \left( {{\cal F} = {{\cal F}_n}} \right)} {\cal{P}}_{\rm{int}}^{{\rm{\cal{Q}}}},
\end{equation}

\noindent where superscript ${\cal{Q}}  \in \left\{\rm {DSC, DSM, DMC, DMM, DMA, DSA} \right\}$ refers to the applicable scheme and  ${\cal{P}}_{\rm{int}}^{{\rm{\cal{Q}}}}=\Pr \left( {C_{{e}}^{{\rm{\cal{Q}}}} > {\cal{R}}} \right)$. In the particular case of interest here, i.e. ${\cal{Q}}=\rm{DMC}$, the following closed-form expression for the IP can be obtained,
\begin{align}
&{\cal{P}}_{\rm{int}}^{{\rm{DMC}}}=\Pr \left( {\frac{1}{2}\rm{lo{g_2}}\left( {1 + {\Psi _{{m^* }e}} + {\Psi _{se}}} \right) > {\cal{R}}} \right) \nonumber\\
&=\! 1\! -\!
 \Pr \!\left( {{\Psi _{{m^* }e}}\! + \!{\Psi _{se}}\! <\! \varrho } \right)\! =\!\! \sum\limits_{l = 1}^5 \!{{\bf{\tilde r}}\left( l \right)\!\exp\! \left( { - {\bf{\tilde t}}\left( l \right)\varrho } \right)},
\end{align}
\noindent where ${\bf{\tilde r}} = \left[ {1, - \rho ,\rho ,\lambda , - \lambda } \right]$, ${\bf{\tilde t}} = \left[ {0,0,\frac{1}{{{{\bar \sigma }_{me}}}},0,\frac{1}{{{{\bar \sigma }_{se}}}}} \right]$, $\rho  = \frac{{{\bar \sigma }_{me}}}{{\left( {{{\bar \sigma }_{me}} - {{\bar \sigma }_{se}}} \right)}}$ and $\lambda  = \frac{{{\bar \sigma }_{se}}}{{\left( {{{\bar \sigma }_{me}} - {{\bar \sigma }_{se}}} \right)}}$.
%
%

Next, we focus on the derivation of the SOP expression. For a DF IR network using $M$ relays, and based on the law of total probability, the following general expression can be obtained for the SOP,
\begin{equation}
{{\cal{P}}_{{\rm{out}}}} = \Pr \left( {{\cal F} = \emptyset } \right){\cal{P}}_{{\rm{out}}}^{{\rm{DT}}}  + \sum\limits_{n = 1}^{{2^M} - 1} {\Pr \left( {{\cal F} = {{\cal F}_n}} \right)} {\cal{P}}_{\rm{out}}^{{\rm{\cal\tilde{Q}}}},
\end{equation}
\noindent where $\cal\tilde{Q} \in \left\{ \rm{DSC,DSM,DSO, DMC, DMM, DMO, DMA, DSA} \right\}$ and ${\cal{P}}_{\rm{out}}^{{\rm{\cal\tilde{Q}}}}=\Pr \left( {C_{{sd}}^{{\rm{\cal\tilde{Q}}}} < {\cal{R}}} \right)$. Note that ${\cal{P}}_{{\rm{out}}}^{{\rm{DT}}}$ was derived in \eqref{Pr_SD_DF}; hence, we next proceed to obtain ${\cal{P}}_{{\rm{out}}}^{{\rm{DMC}}}$. The following lemma and theorem provide key results towards this end.

\begin{lem} \label{CDF-Y}
\emph{The CDF of RV $Y  = \mathop {\max }\limits_{m{\kern 1pt}  \in {\kern 1pt} {{{\cal{F}}_n}}} {\Psi _{md}} + {\Psi _{sd}}$ can be expressed in closed-form as}\\ \vspace{-2mm}
\begin{align}
{F_Y}\left(\varrho  \right) =&\hspace{-.5mm}1 -\hspace{-2mm} \sum\limits_{i = 1}^N \hspace{-1mm}{\frac{{{\pi _{sd}}}}{{{{\bar \sigma }_{id}}\eta }}} \exp \left( { - \frac{\varrho }{{{{\bar \sigma }_{sd}}}}} \right) \hspace{-1mm}-\hspace{-1mm} \sum\limits_{m = 1}^{\left| {{{{\cal{F}}_n}}} \right|}\hspace{-1mm} {\sum\limits_{i = 1}^N {\frac{{{{( - 1)}^{m - 1}}{\pi _{sd}}}}{{{{\bar \sigma }_{sd}}{{\bar \sigma }_{id}}}}}}\times\nonumber\\
&{{\binom{\left| {{{{\cal{F}}_n}}} \right|}{m}} }
\hspace{-1mm}\hspace{-1mm}\left[ \frac{\exp \left( { - \frac{{m\varrho }}{{{{\bar \sigma }_{md}}}}} \right)}{\tilde \eta \left( {\frac{1}{{{{\bar \sigma }_{id}}}} +\frac{{m\varrho }}{{{{\bar \sigma }_{md}}}}} \right)} \hspace{-1mm}+\hspace{-1mm} \frac{\exp \left( { - \frac{\varrho }{{{{\bar \sigma }_{sd}}}}} \right)}{{{\left(\eta\tilde \eta \right) } }} \right],
\label{F_Y}
\end{align}
\noindent where $\tilde \eta   = \left( {\frac{1}{{{{\bar \sigma }_{sd}}}} - \frac{m}{{{{\bar \sigma }_{md}}}}} \right)$ and $\eta  = \frac{\varrho }{{{{\bar \sigma }_{sd}}}} + \frac{1}{{{{\bar \sigma }_{id}}}}$.
\end{lem}
\proof
\emph{See Appendix}~\ref{App-CDF-Y}.
\endproof

The following theorem, whose proof relies on lemma \ref{CDF-Y},  quantifies the SOP for the DMC case.

\begin{theorem} \label{Thm-outage probability-DMC}
\emph{The SOP for the DMC scheme is given by}
\begin{align}
&{\cal{P}}_{\rm{out}}^{{\rm{DMC}}}=
1 - \sum\limits_{l = 1}^2 {\sum\limits_{i = 1}^N {\frac{{{\bf{h}}\left( l \right){\pi _{sd}}}}{{{{\bar \sigma }_{id}}\left( {\varrho  + 1} \right)}}} } \left[ {\chi _{{{\bar \sigma }_{sd}}}} -
\sum_{m = 1}^{\left| {{{{{\cal{F}}_n}}}} \right|}( - 1)^{m - 1}\right.\times\nonumber\\
&\left.\hspace{35mm}{\left| {{{{\cal{F}}_n}}} \right| \choose m}
\frac{{\left[ {{\chi _{{m^{ - 1}}{{\bar \sigma }_{md}}}} - {\chi _{{{\bar \sigma }_{sd}}}}} \right]}}{{{\left( {1 - \frac{{m{{\bar \sigma }_{sd}}}}{{{{\bar \sigma }_{md}}}}} \right)}}} \right],
\label{outage probability-DMC}
\end{align}

\noindent \emph{where} ${\chi _{{{\bar \sigma }_{sd}}}} = {{\bar \sigma }_{sd}}\exp \left( { - \frac{\varrho }{{{{\bar \sigma }_{sd}}}}} \right)\Phi \left( {1,1;\eta  + \frac{{\eta {{\bar \sigma }_{sd}}}}{{\varrho  + 1}}{\bf{g}}\left( l \right)} \right)$ with ${\chi _{{m^{ - 1}}{{\bar \sigma }_{md}}}} $ obtained by using its subscript in place of ${{{\bar \sigma }_{sd}}}$ in ${\chi _{{{\bar \sigma }_{sd}}}}$, ${\bf{h}} = \left[ {\frac{1}{{{{\bar \sigma }_{me}} - {{\bar \sigma }_{se}}}}, - \frac{1}{{{{\bar \sigma }_{me}} - {{\bar \sigma }_{se}}}}} \right]$ and ${\bf{g}} = \left[ {\frac{1}{{{{\bar \sigma }_{me}}}},\frac{1}{{{{\bar \sigma }_{se}}}}} \right]$.
\end{theorem}
\proof
\emph{See Appendix}~\ref{App-DMC-proof}
\endproof

\subsection{Suboptimal Selection Case I: DSC}
In this subsection, we analyze the DF-based IR scheme in which the destination and the eavesdropper node employ SC in order to maximize their respective achievable rate. In this case, the relay that gives the maximum capacity at the destination node is selected, i.e.,
\begin{equation}
{{m^* }=}\mathop {{\rm{arg}}{\rm{max}}}\limits_{m{\kern 1pt}  \in {\kern 1pt} {{\cal{F}}_n}} {\mkern 1mu} \;{C_{md}^{\rm{DF}}},
\end{equation}

\noindent where $C_{md}^{DF}$ is defined in \eqref{eq:C-md}. Then, the secrecy rate for DSC is given by $C_{sd}^{{\rm{DSC}}}= {\left[ {{C_{s}^{\rm{DSC}}} - {C_e^{\rm{DS}}}} \right]^ + } = \frac{1}{2}{\log _2}\left( {\frac{{1 + \max \left\{ {\mathop {\max }\limits_{m\, \in \,{{{\cal{F}}_n}}} {\Psi _{md}},{\Psi _{sd}}} \right\}}}{{1 + \max \left\{ {{\Psi _{{m^* }e}},{\Psi _{se}}} \right\}}}} \right)$,
\noindent where ${C_e^{\rm{DS}}}=\max \left\{ {\frac{1}{2}{{\log }_2}\left( {1 + {\Psi _{me}}} \right),\frac{1}{2}{{\log }_2}\left( {1 + {\Psi _{se}}} \right)} \right\}$, $C_{s}^{{\rm{DSC}}} = \mathop {\max }\limits_{m \in {\cal{F}}_n} C_{sd}^{{\rm{DS}}}$ and $C_{sd}^{{\rm{DS}}} =\max \left\{ {\frac{1}{2}{{\log }_2}\left( {1 + {\Psi _{md}}} \right),\frac{1}{2}{{\log }_2}\left( {1 + {\Psi _{sd}}} \right)} \right\}$. Notice that for the SC technique, the instantaneous SNR is given by the maximum of the two SNRs as in ${C_e^{\rm{DS}}}$ and $C_{sd}^{{\rm{DS}}} $. In the following, we proceed to derive ${\cal{P}}_{\rm{int}}^{{\rm{DSC}}}$, starting with

\begin{equation}
{\cal{P}}_{\rm{int}}^{{\rm{DSC}}}=\Pr \left( {\max \left\{ {{\Psi _{se}},{\Psi _{{m^* }e}}} \right\} > \varrho } \right) = 1-{F_{{\Psi _{{m^* }e}}}}(\varrho ){F_{{\Psi _{se}}}}(\varrho ).
 \end{equation}
Making use of the CDFs of the RVs $\Psi _{se}$ and $\Psi _{me}$, we obtain
 \begin{equation}
{\cal{P}}_{\rm{int}}^{{\rm{DSC}}}= \sum\limits_{l = 1}^3 {{\bf{r}}\left( l \right)\exp \left( { - {\bf{b}}\left( l \right)\varrho } \right)\,\,\,}, \label{Pr-C-e-DS}
\end{equation}
\noindent where \hspace{-4mm}$\quad{\bf{r}} = \left[ {1,1, - 1} \right] {\rm{and}} \; {\bf{b}} = \left[ {\frac{1}{{{{\bar \sigma }_{me}}}},\frac{1}{{{{\bar \sigma }_{se}}}},\frac{1}{{{{\bar \sigma }_{se}}}} + \frac{1}{{{{\bar \sigma }_{me}}}}} \right]$.
\begin{theorem} \label{Thm-outage probability-DSC}
\emph{The SOP for the DSC scheme is given by}
\begin{align}
{\cal{P}}_{\rm{out}}^{{\rm{DSC}}}=& 1 -
\sum\limits_{l = 1}^3 {\sum\limits_{m = 1}^{\left| {{{{{{\cal{F}}_n}}}}} \right|} {\sum\limits_{i = 1}^{{N}} {\frac{{{\bf{a}}\left( l \right){\pi _{sd}}}}{{{{\bar \sigma }_{id}}}}{{( - 1)}^{m - 1}}\binom{\left| {{{{{{\cal{F}}_n}}}}} \right|}{m}} } }\times\nonumber\\
 &\left[ {{{\cal{I}}_{{m^{ - 1}}{{\bar \sigma }_{md}}}} - {{\cal{I}}_{{\tau ^{ - 1}}}}} \right]- \sum\limits_{l = 1}^3 {\sum\limits_{i = 1}^{{N}} {\frac{{{\bf{a}}\left( l \right){\pi _{sd}}}}{{{{\bar \sigma }_{id}}}}} } {{\cal{I}}_{{{\bar \sigma }_{sd}}}},
\label{outage probability-DSC}
\end{align}
\noindent where ${{\cal{I}}_{{{\bar \sigma }_{sd}}}} = \exp \left( { - \frac{\varrho }{{{{\bar \sigma }_{sd}}}}} \right)\frac{{{{\bar \sigma }_{sd}}}}{{\varrho  + 1}}{\rm{\Phi }}\left( {1,1;\eta  + {\bf{b}}\left( l \right)\frac{{{{\bar \sigma }_{sd}}\eta }}{{\varrho  + 1}}} \right)$, $\tau  = \left( {\frac{m}{{{{\bar \sigma }_{md}}}} + \frac{1}{{{{\bar \sigma }_{sd}}}}} \right)$ and ${\bf{a}} = \left[ {\frac{1}{{{{\bar \sigma }_{me}}}},\frac{1}{{{{\bar \sigma }_{se}}}}, - \left( {\frac{1}{{{{\bar \sigma }_{se}}}} + \frac{1}{{{{\bar \sigma }_{me}}}}} \right)} \right]$.

\begin{proof}
\emph{The proof follows the same steps as that of Theorem}~\ref{Thm-outage probability-DMC}.
\end{proof}
\end{theorem}

\subsection{Suboptimal Selection Case II: DMM}
We now investigate the use of MRC with the minimum selection scheme (DMM) under Case II, where additional CSI information about the $R_m\hspace{-1mm}-\hspace{-1mm}E$ links is available. The objective is to select the relay in ${\cal{F}}_n$ to minimize the achievable rate at the eavesdropper node. This relay selection scheme considers only the $R_m\hspace{-1mm}-\hspace{-1mm}E$ link and furthermore, both the destination and eavesdropper nodes employ MRC. In this case, the relay that yields the lowest instantaneous rate at the eavesdropper will be selected, i.e.,
\begin{equation}
{{m^* }= } \mathop {{\rm{arg}}{\rm{min}}}\limits_{m{\kern 1pt}  \in {\kern 1pt} {{\cal{F}}_n}} {\mkern 1mu}\; {C_{e}^{\rm{DMM}}}.
\end{equation}
Consequently, the secrecy rate becomes $C_{sd{\rm{}}}^{{\rm{DMM}}} = {\left[ {{C_{sd}^{\rm{DM}}} -  {C_e^{\rm{DMM}}}} \right]^ + } = \frac{1}{2}{\log _2}\left( {\frac{{1 + {{\Psi _{{m^* }d}}+{\Psi _{sd}}} }}{{1 +  {\mathop {\min }\limits_{m\, \in \,{{\cal{F}}_n}} {\Psi _{me}}+{\Psi _{se}}} }}} \right)$, where ${C_e^{\rm{DMM}}}=\mathop {{\rm{}}{\rm{min}}}\limits_{m{\kern 1pt}  \in {\kern 1pt} {{\cal{F}}_n}}{C_e^{\rm{DM}}}$. The IP expression for this case can be obtained as
\begin{align}
&{\cal{P}}_{\rm{int}}^{{\rm{DMM}}}= \Pr \left( {\frac{1}{2}{\rm{lo{g}}_2}\left( {1 + \mathop {\min }\limits_m \,{\Psi _{me}} + {\Psi _{se}}} \right) > {\cal{R}}} \right)\nonumber\\&
  = 1 -\Pr \left( {\mathop {\min }\limits_m \,{\Psi _{me}} + {\Psi _{se}} < \varrho } \right)
\hspace{-1mm} \nonumber\\&
= 1\hspace{-1mm} -\hspace{-1mm} \sum\limits_{l = 1}^2 {\frac{{{\bf{\tilde h}}\left( l \right)}}{{{\bf{\tilde g}}\left( l \right)}}\left[ {1 - \exp \left( { - {\bf{\tilde g}}\left( l \right)\varrho } \right)} \right]},
\end{align}

\noindent where ${\bf{\tilde g}}\left( l \right)= \left[ {\frac{1}{{{{\bar \sigma }_{se}}}},\frac{{\left| {{{{\cal{F}}_n}}} \right|}}{{{{\bar \sigma }_{me}}}}} \right]$ and ${\bf{\tilde h}}\left( l \right) = \left[ {\frac{{\left| {{{{\cal{F}}_n}}} \right|}}{{\left( {\left| {{{{\cal{F}}_n}}} \right|{{\bar \sigma }_{se}} - {{\bar \sigma }_{me}}} \right)}}, - \frac{{\left| {{{{\cal{F}}_n}}} \right|}}{{\left( {\left| {{{{\cal{F}}_n}}} \right|{{\bar \sigma }_{se}} - {{\bar \sigma }_{me}}} \right)}}} \right]$. \\

 To proceed with the derivation of the SOP, we first need to obtain a closed-form expression for the CDF of the RV $\tilde Y ={{\Psi _{{m^* }d}}+{\Psi _{sd}}} $, which is presented in the following lemma.

\begin{lem} \label{CDF-X}
\emph{The CDF of $\tilde Y $ is given by}
\begin{align}
{F_{\tilde Y}}(\gamma ) =& 1 - \sum\limits_{i = 1}^N  \frac{\pi _{sd}}{ 1+{\frac{{\bar \sigma }_{id} }{{{{\bar \sigma }_{sd}}}}\gamma }}\left[ {\exp \left( { - \frac{\gamma }{{{{\bar \sigma }_{sd}}}}} \right)}
 { \hspace{-1mm}+\hspace{-.2mm} \frac{{\exp \left( { - \frac{\gamma }{{{{\bar \sigma }_{sd}}}}} \right)}}{{{{\bar \sigma }_{sd}}\left( {\frac{1}{{{{\bar \sigma }_{sd}}}} \hspace{-1mm}-\hspace{-1mm} \frac{1}{{{{\bar \sigma }_{md}}}}} \right)}}} \right]\nonumber\\&
  - \frac{{\exp \left( { - \frac{\gamma }{{{{\bar \sigma }_{md}}}}} \right)}}{{{{\bar \sigma }_{sd}}\left( {\frac{1}{{{{\bar \sigma }_{sd}}}} \hspace{-1mm}- \hspace{-1mm}\frac{1}{{{{\bar \sigma }_{md}}}}} \right)}}\sum\limits_{i = 1}^N {\frac{{{\pi _{sd}}}}{{{{\bar \sigma }_{id}}}}} \frac{1}{{\left( {\frac{\gamma }{{{{\bar \sigma }_{md}}}} \hspace{-1mm}+\hspace{-1mm} \frac{1}{{{{\bar \sigma }_{id}}}}} \right)}}.
\label{F_Y_tilda}
\end{align}
\begin{proof}
\emph{The proof of this lemma is analogous to that of Lemma}~\ref{CDF-Y}.
\end{proof}
\end{lem}

We are now in a position to derive the desired SOP expression, which is provided in the following theorem.

\begin{theorem} \label{Thm-outage probability-DMM}
\emph{The SOP for the DMM scheme is given by}
\begin{equation}
{\cal{P}}_{\rm{out}}^{{\rm{DMM}}} \! =\! 1 \!-\! \sum\limits_{l = 1}^2 \sum\limits_{m = 1}^{\left| {{{{{{\cal{F}}_n}}}}} \right|}{\sum\limits_{i = 1}^{{N}} {\frac{{{\bf{\tilde h}(l)}{\pi _{sd}}}\left[ {{{\cal{T}}_{sd}} + \vartheta {{\cal{T}}_{md}} - \vartheta {{\cal{T}}_{sd}}} \right]}{{{{\bar \sigma }_{id}}\left( {\varrho  + 1} \right)}}} } ,
\label{outage probability-DMM}
\end{equation}

\noindent \emph{where} ${{\cal{T}}_{sd}} = {{\bar \sigma }_{sd}}\exp \left( { - \frac{\varrho }{{{{\bar \sigma }_{sd}}}}} \right)\Phi \left( {1,1;\eta  + \frac{{{{\bf{\tilde g}}\left( l \right)}\eta {{\bar \sigma }_{sd}}}}{{\varrho  + 1}}} \right)$ \emph{and} $\vartheta  = \left( {1 - \frac{{{{\bar \sigma }_{sd}}}}{{{{\bar \sigma }_{md}}}}} \right)$.
\begin{proof}
\emph{See Appendix}~\ref{App-DMMM-proof}.
\end{proof}
\end{theorem}
%
%
\subsection{Suboptimal Selection Case II: DSM}
For this case, the relay is chosen according to the following rule,
\begin{equation}
{{m^* }= } \mathop {{\rm{arg min}}}\limits_{m{\kern 1pt}  \in {\kern 1pt} {\cal{F}}_n} \;{C_e^{\rm{DS}}},
\end{equation}
\noindent while the secrecy rate is given by $C_{sd{\rm{}}}^{{\rm{DSM}}} ={\left[ {{C_{sd}^{\rm{DS}}} -  {C_e^{\rm{DSM}}}} \right]^ + } = \frac{1}{2}{\log _2}\left( {\frac{{1 + \max \left\{ {{\Psi _{{m^* }d}},{\Psi _{sd}}} \right\}}}{{1 + \max \left\{ {\mathop {\min }\limits_{m\, \in \,{{\cal{F}}_n}} {\Psi _{me}},{\Psi _{se}}} \right\}}}} \right)$, where $ {C_e^{\rm{DSM}}}=\mathop {{\rm{min}}}\limits_{m \in {\cal{F}}_n} {C_e^{\rm{DS}}}$.

Herein we define the variable $u  = \mathop {\min }\limits_{m \in  {{{{\cal{F}}_n}}}} {\Psi _{me}}$ with CDF ${F_U}(\gamma ) = 1 - \exp \left( { - \frac{{\left| {{{{\cal{F}}_n}}} \right|\gamma }}{{{{\bar \sigma }_{me}}}}} \right)$, in terms of which the intercept probability for the DSM case can be expressed as
\begin{align}
{\cal{P}}_{\rm{int}}^{{\rm{DSM}}}&= \Pr \left( {\frac{1}{2}\rm{lo{g_2}}\left( {1 + \max \left\{  {u,{\Psi _{se}}} \right\}} \right) > {\cal{R}}} \right) \hspace{-1mm}
\nonumber\\
&=\hspace{-1mm} 1 \hspace{-1mm}-\hspace{-1mm} {F_u}(\varrho ){F_{{\Psi _{se}}}}(\varrho )
\hspace{-1mm}
 = \hspace{-1mm}\sum\limits_{l = 1}^3 {{\bf{r}}\left( l \right)\exp \left( { - {\bf{\tilde b}}\left( l \right)w} \right)},
 \label{Pr-C-e-DSM}
\end{align}

\noindent where ${\bf{\tilde b}} = \left[ {\frac{{\left| {{{{\cal{F}}_n}}} \right|}}{{{{\bar \sigma }_{me}}}},\frac{1}{{{{\bar \sigma }_{se}}}},\left( {\frac{1}{{{{\bar \sigma }_{se}}}} + \frac{{\left| {{{{\cal{F}}_n}}} \right|}}{{{{\bar \sigma }_{me}}}}} \right)} \right]$.

Besides, the SOP can be obtained in closed-form as given in the following theorem.
\begin{theorem} \label{Thm-outage probability-DSM}
\emph{The SOP for the DSM scheme is given by}
\begin{equation}
{\cal{P}}_{\rm{out}}^{{\rm{DSM}}} \!=\! 1\! -\! \sum\limits_{l = 1}^3 \sum\limits_{m = 1}^{\left| {{{{{{\cal{F}}_n}}}}} \right|}{\sum\limits_{i = 1}^{{N}} {\frac{{{\bf{\tilde a}}\left( l \right){\pi _{sd}}}\left[ {{{\cal{J}}_{{{\bar \sigma }_{md}}}} \!+\! {{\cal{J}}_{{{\bar \sigma }_{sd}}}}\! -\! {{\cal{J}}_{{{\tilde \tau }^{ - 1}}}}} \right]}{{{{\bar \sigma }_{id}}(\varrho  + 1)}}}},
\label{outage probability-DSM}
\end{equation}

\noindent where ${{\cal{J}}_{{{\bar \sigma }_{md}}}} = {{\bar \sigma }_{md}}\exp \left( { - \frac{\varrho }{{{{\bar \sigma }_{md}}}}} \right)\Phi \left( {1,1;\mu  + \frac{{\mu {{\bar \sigma }_{md}}}}{{(\varrho  + 1)}}{\bf{\tilde b}}\left( l \right)} \right)$, ${\bf{\tilde a}} = \left[ {\frac{{\left| {{{{\cal{F}}_n}}} \right|}}{{{{\bar \sigma }_{me}}}},\frac{1}{{{{\bar \sigma }_{se}}}}, - \left( {\frac{1}{{{{\bar \sigma }_{se}}}} + \frac{{\left| {{{{\cal{F}}_n}}} \right|}}{{{{\bar \sigma }_{me}}}}} \right)} \right]$ and $\tilde \tau  = \left( {\frac{1}{{{{\bar \sigma }_{md}}}} + \frac{1}{{{{\bar \sigma }_{sd}}}}} \right)$.

\begin{proof}
\emph{The proof is similar to that of Theorem}~\ref{Thm-outage probability-DMM}.
\end{proof}
\end{theorem}

\subsection{Optimal Selection Case III: DSO}
The two previously considered relay selection schemes do not simultaneously involve the relay to destination and relay to eavesdropper channels. In contrast, the optimal relay selection scheme takes into account CSI information for both the mentioned links. This subsection presents the DSO scheme where the relay selected for forwarding the source signal to the destination is the one achieving the maximum secrecy capacity, which by definition takes into account the quality of both links.  Specifically, the desired relay is chosen as
\begin{equation}
{{m^* =} } \mathop {{\rm{arg max}}}\limits_{m{\kern 1pt}  \in {\kern 1pt} {\cal{F}}_n}\; C_{m{\rm{}}}^{{\rm{b}}},
\label{asas}
\end{equation}
\noindent where $C_m^b  = \frac{1}{2}{\log _2}\left( { \frac{{1 + {\Psi _{md}}}}{{1 + {\Psi _{me}}}}} \right)$. The secrecy rate for this case is given by~\citep{Shen:TVT:2014}
\begin{equation}
C_{sd{\rm{}}}^{{\rm{DSO}}}  =  {\max \big\{ {{C^a},\mathop {\max }\limits_{m{\kern 1pt}  \in {\kern 1pt} {{{\cal{F}}_n}}} C_m^b} \big\}}. \label{C-sd-DSo}
\end{equation}
\noindent where ${C^a}  = \frac{1}{2}{\log _2}\left( {\frac{{1 + {\Psi _{sd}}}}{{1 + {\Psi _{se}}}}} \right)$. We notice that the derivation of the SOP in this case is quite challenging and it does not seem possible to obtain a closed-form expression. Therefore, we rely on the approximation of the SOP at high SNR in our study, as further developed in Section~\ref{DO-Sec}.
%
%
%
%
%
\subsection{Optimal Selection Case III: DMO}
In the case of DMO, the proposed selection technique selects the optimal relay as in \eqref{asas}, and the secrecy rate will be
%
%
\begin{equation}
C_{sd{\rm{}}}^{{\rm{DMO}}}  = \frac{1}{2}{\log _2}\left( {{2^{2{C^a}}} + \mathop {\max }\limits_{m \in {{{\cal{F}}_n}} } {2^{2C_m^b}}} \right).\label{DMO-op}
\end{equation}

Likewise the DSO case, obtaining a closed-form expression for the SOP in the DMO case is intractable. However, a closed form expression for the SOP in the high SNR regime will be obtained in Section~\ref{DO-Sec}. Nevertheless, numerical SOP results for the DSO and DMO cases can be obtained through computer simulations.

\subsection{Suboptimal Selection: DSA}
Thus far, emphasis has been placed on the cases in which only the best relay was employed in the cooperation phase. The DSA scheme considers the case where several relays (i.e. more than 1) can assist in forwarding confidential information from $S$ to $D$. To be specific, all the relays in the WIRS re-encode the information and forward this re-encoded message to the destination (and eavesdropper). This subsection assumes that both the destination and the eavesdropper use SC technique. Hence, the secrecy rate is defined as
\begin{equation}
\hspace{-2mm}C_{sd}^{{\rm{DSA}}} \!\hspace{-1mm}=\hspace{-1mm} {\left[ {C_{md}^{{\rm{DSA}}} \hspace{-1mm}-\hspace{-1mm} C_{me}^{{\rm{DSA}}}} \right]^ + } \! \hspace{-1mm}=\hspace{-1mm} \frac{1}{2}{\rm{log}}\!\left( \!{\frac{{1 \hspace{-1mm}+\hspace{-1mm} \max\! \left\{\! {\Psi _{md}^{{\rm{DSA}}}\! , {\Psi _{sd}}} \right\}}}{{1 \hspace{-1mm}+\hspace{-1mm} \max \!\left\{ \!{\Psi _{me}^{{\rm{DSA}}}\! , {\Psi _{se}}} \right\}}}}\!\! \right),\label{C-DSA-F}
\end{equation}
\noindent where $P'_m=P'/\left(\left| {{\cal{F}}_n} \right|+1\right)$, $\Psi _{md}^{{\rm{DSA}}} = \frac{{\mathop {\max }\limits_{{m \in \cal{F}}_n} {P'_m}{{\left| {{h_{md}}} \right|}^2}}}{{\sum\limits_{i = 1}^{{N}} {{P_{i}}{{\left| {{c_{id}}} \right|}^2}}  + \sigma _n^2}}$ and $\Psi _{me}^{{\rm{DSA}}} = \mathop {\max }\limits_{{m \in \cal{F}}_n} \frac{{{P'_m}{{\left| {{h_{m,e}}} \right|}^2}}}{{\sigma _n^2}}$.

With the assumption that both the destination and the eavesdropper node employ SC, the IP of the DSA scheme can be formulated as
\begin{equation}
{\cal{P}}_{{\rm{int}}}^{{\rm{DSA}}}\hspace{-1mm}=\hspace{-1.5mm} \sum\limits_{m = 1}^{\left| {{{{\cal{F}}_n}}} \right|} \hspace{-1mm}{\frac{{{{( - 1)}^{m \hspace{-.3mm}-\hspace{-.3mm} 1}}}}{{{{\bar \sigma }_{se}}\hat \omega }}\binom{\left| {{{{\cal{F}}_n}}} \right|}{m}} \hspace{-1mm}\left[ {\exp \hspace{-1mm}\left(\hspace{-1mm} { - \frac{{m\varrho }}{{{{\bar \sigma }_{me}}}}}\hspace{-1mm} \right)\hspace{-1mm} -\hspace{-1mm} \exp\hspace{-1mm} \left(\hspace{-1mm} { - \frac{\varrho }{{{{\bar \sigma }_{se}}}} \hspace{-1mm}-\hspace{-1mm} \hat \omega \varrho } \right)}\hspace{-1mm} \right]\hspace{-1mm}.
\end{equation}

\noindent where $\hat \omega   = \left( {\frac{1}{{{{\bar \sigma }_{se}}}} - \frac{m}{{{{\bar \sigma }_{me}}}}} \right)$. We next develop a closed-form expression of the secrecy outage performance for the DSA scheme. To begin, we first introduce the following key result. 

\begin{lem} \label{Thm-PDF-DSA}
\emph{Let the denominator of the log function in \eqref{C-DSA-F} be $1+\gamma_1$. Then, the pdf of $\gamma_1$ is derived as}
\begin{align}
{f_{{\gamma _1}}}\left( \gamma  \right) = \sum\limits_{{m_1} = 1}^{\left| {{{\cal{F}}_n}} \right|} {\sum\limits_{l = 1}^4 {{{( - 1)}^{{m_1} - 1}}} \left( \begin{array}{l}
\left| {{F_n}} \right|\\
{m_1}
\end{array} \right){\bf{\tilde c}}\left( l \right)\exp \left( { - {\bf{c}}\left( l \right)\gamma } \right)}
\label{PDF-Psi-md-DSA}
\end{align}
\noindent where ${\bf{\tilde c}} = \left[ {\frac{{{m_1}}}{{{{\bar \sigma }_{me}}}}, - \frac{{{m_1}}}{{{{\bar \sigma }_{me}}}}, - \frac{1}{{{{\bar \sigma }_{se}}}},\frac{1}{{{{\bar \sigma }_{se}}}}} \right],{\bf{c}} = \left[ {\frac{{{m_1}}}{{{{\bar \sigma }_{me}}}},\hat \varpi ,\hat \varpi ,\frac{1}{{{{\bar \sigma }_{se}}}}} \right]$ and $\hat \varpi  = \left( {\frac{{{m_1}}}{{{{\bar \sigma }_{me}}}} + \frac{1}{{{{\bar \sigma }_{se}}}}} \right)$.
\end{lem}
\proof
The CDF of ${\Psi _{me}^{{\rm{DSA}}}}$ and ${{\Psi _{se}}}$ is obtained respectively as ${F_{\Psi _{me}^{{\rm{DSA}}}}}\left( \gamma  \right) = 1 - \sum\limits_{m = 1}^{\left| {{{\cal{F}}_n}} \right|} {{{( - 1)}^{m - 1}}\left( \begin{array}{l}
\left| {{{\cal{F}}_n}} \right|\\
m
\end{array} \right)\exp \left( { - \frac{{m\gamma }}{{{{\bar \sigma }_{me}}}}} \right)}$ and
${F_{{\Psi _{se}}}}\left( \gamma  \right) = \left[ {1 - \exp \left( { - \frac{\gamma }{{{{\bar \sigma }_{se}}}}} \right)} \right]$. Then, by taking the derivative of ${f_{{\gamma _1}}}\left( \gamma  \right) =\frac{d}{{d\gamma }}\left[ {{F_{\Psi _{mE}^{{\rm{DSA}}}}}\left( \gamma  \right) \times {F_{{\Psi _{se}}}}\left( \gamma  \right)} \right]$ we obtain \eqref{PDF-Psi-md-DSA}.\\
\endproof

Lemma~\ref{Thm-PDF-DSA} allows us to obtain a closed-form expression for the secrecy rate of the DSA scheme as stated in the following theorem.

\begin{theorem} \label{Thm-outage probability-DSA}
\emph{The SOP for the DSA scheme is given by}
\begin{align}
{\cal{P}}_{\rm{out}}^{{\rm{DSA}}}& =1 - \left[ {{I_{{{\bar \sigma }_{sd}}}} + \sum\limits_{m = 1}^{\left| {{F_n}} \right|} {{{( - 1)}^{m - 1}}\left( \begin{array}{l}
\left| {{{\cal{F}}_n}} \right|\\
m
\end{array} \right)\left[ {{I_{{{\bar \sigma }_{md}}{m^{ - 1}}}} - {I_{{\tau ^{ - 1}}}}} \right]} } \right],
  \label{outage probability-DSA}
\end{align}
\emph{where}
\begin{align}
{I_{{{\bar \sigma }_{sd}}}} =& \sum\limits_{i = 1}^N {\sum\limits_{{m_1} = 1}^{\left| {{{\cal{F}}_n}} \right|} {\sum\limits_{l = 1}^4 {\frac{{{\pi _{id}}}}{{{{\bar \sigma }_{id}}}}} {{( - 1)}^{{m_1} - 1}}{\bf{\tilde c}}\left( l \right){{\left( {\frac{{{2^{2R}}}}{{{{\bar \sigma }_{sd}}}}} \right)}^{ - 1}}\left( \begin{array}{l}
\left| {{{\cal{F}}_n}} \right|\\
{m_1}
\end{array} \right)} }  \times \notag\\&
\exp \left( { - \frac{\varrho }{{{{\bar \sigma }_{sd}}}}} \right)\Phi \left( {1,1;{\bf{c}}\left( l \right)\eta \frac{{{{\bar \sigma }_{sd}}}}{{{2^{2R}}}} + \eta } \right),
\end{align}
 and $\eta  = \left( {\frac{1}{{{{\bar \sigma }_{id}}}} + \frac{\varrho }{{{{\bar \sigma }_{sd}}}}} \right)$.
\end{theorem}
\proof
Let the numerator of the log function in \eqref{C-DSA-F} be $1+\gamma_2$. Using the PDF of RV $\gamma_1$ as in \eqref{PDF-Psi-md-DSA} as well as the CDF of RV $Y$ in \eqref{F_Y}, we express the secrecy rate of the DSA scheme as
\begin{align}
{\cal{P}}_{\rm{out}}^{{\rm{DSA}}}&={{\rm{E}}_{{\gamma _1}}}\left[ {\Pr \left( {{\gamma _2} < {2^{2R}}{\gamma _1} + \varrho } \right)} \right]\\&
=\int_0^\infty  {{F_{{\gamma _2}}}} \left( {{2^{2R}}{\gamma _1} +\varrho } \right){f_{{\gamma _1}}}\left( {{\gamma _1}} \right)d{\gamma _1}.
\label{dsfs}
\end{align}
The desired result is obtained by substituting \eqref{PDF-Psi-md-DSA} into \eqref{dsfs} and evaluating the resulting integral.
\endproof

\vspace{-4mm}

\subsection{Suboptimal Selection: DMA}
This scheme is analogous to the DSA one except that the destination and the eavesdropper both employ the MRC technique. In this case, the secrecy rate is
\begin{equation}
C_{sd}^{{\rm{DMA}}} = {\left[ {C_{md}^{{\rm{DMA}}} - C_{me}^{{\rm{DMA}}}} \right]^ + } = \frac{1}{2}{\rm{log}}\left( {\frac{{1 + \Psi _{md}^{{\rm{DMA}}}  {}}}{{1 + \Psi _{me}^{{\rm{DMA}}}  {}}}} \right),
\end{equation}

\noindent where $\Psi _{md}^{{\rm{DMA}}} = \frac{{\sum\limits_{m \in {{{\cal{F}}_n}}}^{} {{P'_m}{{\left| {{h_{md}}} \right|}^2}} }}{{\sum\limits_{i = 1}^{{N}} {{P_{id}}{{\left| {{c_{id}}} \right|}^2}}  + \sigma _n^2}}+{\Psi _{sd}}$ and $\Psi _{me}^{{\rm{DMA}}} = \sum\limits_{m \in {{{\cal{F}}_n}}}^{} {\frac{{{P'_m}{{\left| {{h_{m,E}}} \right|}^2}}}{{\sigma _n^2}}}+\Psi _{se} $.

In the following, we analyze the IP of the DMA case in which all relays that can successfully decode the source's message simultaneously forward its replicated image to the destination. For this DMA case, the IP is defined as
\begin{equation}
{\cal{P}}_{{\rm{int}}}^{{\rm{DMA}}} = \Pr \left( {\frac{1}{2}{{\log }_2}\left( {1 + \Psi _{me}^{{\rm{DMA}}} {}} \right) > {\cal{R}}} \right).
\end{equation}
The latter can be expressed in closed form as
\begin{align}
{\cal{P}}_{{\rm{int}}}^{{\rm{DMA}}}=&\sum\limits_{k = 0}^{\left| {{{\cal{F}}_n}} \right| - 1} {\sum\limits_{t = 0}^k {\frac{{{\varrho ^{k - t}}{{\left( { - 1} \right)}^t}}}{{{{\left( {{{\bar \sigma }_{me}}} \right)}^k}{{\bar \sigma }_{se}}k!}}\binom{k}{t}} }\times\nonumber\\
&\left[ {\frac{{t!}}{{{\zeta ^{t + 1}}}} - \exp \left( { - \varrho \zeta } \right)\sum\limits_{i = 0}^t {\frac{{t!{\varrho ^i}}}{{i!{\zeta ^{t - i + 1}}}}} } \right],
\end{align}
\noindent where $\zeta   = \left( {\frac{1}{{{{\bar \sigma }_{se}}}} - \frac{1}{{{{\bar \sigma }_{me}}}}} \right)$.\\

Next, we proceed to obtain the SOP of the DMA scheme which can be expressed as
\begin{equation}
\hspace{-3mm}\Pr \left( {C_{sd}^{{\rm{DMA}}}\! <\! {\cal{R}}} \!\right)  \hspace{-1mm}=\hspace{-1mm} {{\rm{E}}_{\Psi _{me}^{{\rm{DMA}}}}}\!\left[ {{F_{\Psi _{md}^{{\rm{DMA}}}}}\!\left( {\varrho  + \left( {\varrho  + 1} \right)\Psi _{me}^{{\rm{DMA}}}} \right)} \right]\!.
\label{OP-DMA-D}
\end{equation}

In order to proceed with the evaluation of  \eqref{OP-DMA-D}, we first need to obtain a closed-from expression for the CDF of ${\Psi _{md}^{{\rm{DMA}}}}$, which is provided in the following lemma.
\begin{lem} \label{Lem-CDF-Psi-md-DMA}
\emph{The CDF of ${\Psi _{md}^{{\rm{DMA}}}}$ is derived as}\vspace{-2mm}
\begin{align}
&{F_{\Psi _{md}^{{\rm{DMA}}}}}\left( \varrho  \right) \hspace{-1mm}=\hspace{-1mm} 1- \sum\limits_{m = 1}^{\left| {{{\cal{F}}_n}} \right|} {\sum\limits_{k = 0}^{m - 1} {\sum\limits_{l = 0}^k {\sum\limits_{i = 1}^N {\frac{{{\upsilon _m}{\pi _{sd}}\bar \sigma _{md}^{m - k}{\varrho ^k}\Gamma \left( {l \hspace{-1mm}+\hspace{-1mm} 1} \right)}}{{{{\bar \sigma }_{id}}\Gamma \left( {k + 1} \right)}}}}}}\times\nonumber\\
&{{{{\binom{k}{l}} } } } \,\frac{{\exp \left( { - \frac{\varrho }{{{{\bar \sigma }_{md}}}}} \right)}}{{{\Xi ^{l + 1}}}}- \sum\limits_{i = 1}^N {\frac{{\omega {{\bar \sigma }_{sd}}{\pi _{sd}}}}{{{{\bar \sigma }_{id}}\eta }}} \exp \left( { - \frac{\varrho }{{{{\bar \sigma }_{sd}}}}} \right),
\label{CDF-Psi-md-DMA}
\end{align}
\emph{where} $\Xi  = \left( {\frac{\varrho }{{{{\bar \sigma }_{md}}}} + \frac{1}{{{{\bar \sigma }_{id}}}}} \right)$, $\omega  = \frac{1}{{{{\bar \sigma }_{sd}}\bar \sigma _{md}^{\left| {{{\cal{F}}_n}} \right|}}}\frac{1}{{{{\left( {\frac{1}{{{{\bar \sigma }_{md}}}} - \frac{1}{{{{\bar \sigma }_{sd}}}}} \right)}^{\left| {{{\cal{F}}_n}} \right|}}}}$ and ${\upsilon _m} = \frac{{{{\left( { - 1} \right)}^{\left| {{{\cal{F}}_n}} \right| - m}}}}{{{{\bar \sigma }_{sd}}\bar \sigma _{md}^{\left| {{{\cal{F}}_n}} \right|}}}\frac{1}{{{{\left( {\frac{1}{{{{\bar \sigma }_{sd}}}} - \frac{1}{{{{\bar \sigma }_{md}}}}} \right)}^{\left| {{{\cal{F}}_n}} \right| - m + 1}}}}$.
\end{lem}
\proof
\emph{See Appendix~\ref{App-lem-DMM}}
\endproof


Now, with the help of Lemma~\ref{Lem-CDF-Psi-md-DMA}, the final SOP expression for the DMA case can be obtained, as stated in the following theorem.

\begin{theorem} \label{Thm-outage probability-DMA}
\emph{The SOP for the DMA scheme is given by \eqref{outage probability-DMA} shown on the top of the next page,}
\begin{figure*}
\small
\begin{align}
\Pr \left( {C_{{sd}}^{{\rm{DMA}}} < {\cal{R}}} \right) =&
1 - \sum\limits_{i = 1}^N {\frac{{\omega \bar \sigma _{sd}^2{\pi _{sd}}}}{{{{\bar \sigma }_{id}}\left( {\varrho  + 1} \right)}}} \exp \left( { - \frac{\varrho }{{{{\bar \sigma }_{sd}}}}} \right) \times \nonumber\\&
\left[ {q'\varphi \left( {1,1;\alpha \left( {1 + \frac{{{{\bar \sigma }_{sd}}}}{{{{\bar \sigma }_{se}}\left( {\varrho  + 1} \right)}}} \right)} \right) + \sum\limits_{q = 1}^{\left| {{{\cal{F}}_n}} \right|} {{\xi' _q}{{\left( {\frac{{\alpha {{\bar \sigma }_{sd}}}}{{\varrho  + 1}}} \right)}^{q - 1}}} \Phi \left( {q,q;\alpha '\left( {1 + \frac{{{{\bar \sigma }_{sd}}}}{{{{\bar \sigma }_{me}}\left( {\varrho  + 1} \right)}}} \right)} \right)} \right]\nonumber\\&
 - \sum\limits_{m = 1}^{\left| {{{\cal{F}}_n}} \right|} {\sum\limits_{k = 0}^{i - 1} {\sum\limits_{L = 0}^k {\sum\limits_{i = 1}^N {\sum\limits_{p = 0}^k {\frac{{{\upsilon _m}{\pi _{sd}}\bar \sigma _{md}^{L + m - k + 1}q' {\varrho ^{k - p}}\Gamma \left( {L + 1} \right)}}{{{{\bar \sigma }_{id}}{{\left( {\varrho  + 1} \right)}^{L - p + 1}}\Gamma \left( {k + 1} \right)}}\left( \begin{array}{l}
k\\
p
\end{array} \right)} \left( \begin{array}{l}
k\\
L
\end{array} \right)} } } } \exp \left( { - \frac{\varrho }{{{{\bar \sigma }_{md}}}}} \right) \times \nonumber\\&
\left[ {\Gamma \left( {p + 1} \right){{\left( {\frac{{\beta {{\bar \sigma }_{md}}}}{{\varrho  + 1}}} \right)}^{p - L}}\,\Phi \left( {p + 1,p - L + 1;\beta \left( {1 + \frac{{{{\bar \sigma }_{md}}}}{{{{\bar \sigma }_{se}}\left( {\varrho  + 1} \right)}}} \right)} \right)} \right.\nonumber\\&
 + \left. {\sum\limits_{q = 1}^{\left| {{{\cal{F}}_n}} \right|} {\frac{{{\xi' _q}\Gamma \left( {p + q} \right)}}{{\Gamma \left( q \right)}}{{\left( {\frac{{{{\bar \sigma }_{md}}\beta }}{{\varrho  + 1}}} \right)}^{p + q - L - 1}}} \Phi \left( {p + q,p + q - L;\beta \left( {1 + \frac{{{{\bar \sigma }_{md}}}}{{{{\bar \sigma }_{me}}\left( {\varrho  + 1} \right)}}} \right)} \right)} \right].
 \label{outage probability-DMA}
\end{align}
\hrule
\end{figure*}
\normalsize
\noindent \emph{where} $q'=\frac{1}{{{{\bar \sigma }_{se}}\bar \sigma _{me}^{\left| {{{\cal{F}}_n}} \right|}}}\frac{1}{{{{\left( {\frac{1}{{{{\bar \sigma }_{me}}}} - \frac{1}{{{{\bar \sigma }_{se}}}}} \right)}^{\left| {{{\cal{F}}_n}} \right|}}}}$ and ${\xi' _q} = \frac{{{{\left( { - 1} \right)}^{\left| {{{\cal{F}}_n}} \right| - m}}}}{{{{\bar \sigma }_{se}}\bar \sigma _{me}^{\left| {{{\cal{F}}_n}} \right|}}}\frac{1}{{{{\left( {\frac{1}{{{{\bar \sigma }_{se}}}} - \frac{1}{{{{\bar \sigma }_{me}}}}} \right)}^{\left| {{{\cal{F}}_n}} \right| - m + 1}}}}$.
\end{theorem}
\proof
See Appendix~\ref{Apx-Thm-outage probability-DMA}
\endproof



\section{Diversity Order Analysis}\label{DO-Sec}
In this section, to characterize the impact of key parameters on the secrecy outage performance, the asymptotic SOP in the high SNR regime is
investigated. To simplify the developments, we let  ${{\bar \sigma }_{sd}} =\varepsilon {{\bar \sigma }_{md}} $ and ${{\bar \sigma }_{sm}} =\hat \varepsilon {{\bar \sigma }_{md}} $, where $\varepsilon$ and $\hat \varepsilon$ are positive numbers close to $1$, which means that the channel quality of the legitimate links is comparable (of the same order). We also let ${{\bar \sigma }_{se}} = \tilde \varepsilon {{\bar \sigma }_{me}}$ with $\tilde \varepsilon$ close to 1, meaning that the channel quality of the wiretap links is similar. We first consider the case ${{\bar \sigma }_{sd}} \to \infty $, which corresponds to a scenario where $S$ is located much closer to $D$ than $ E$. Subsequently, we also consider the limiting case where ${{\bar \sigma }_{se}} \to \infty $, for which the intercept probability goes to $1$. Below, we first derive SOP expressions in the asymptotic regime for each one of the following relay selection schemes: DSA, DSM, DSC, DMA, DMC, DMM, DSO and DMO. Using these formulas, we then derive corresponding expressions of the coding gain and diversity.

\subsection{Analysis}
\noindent \emph{Direct Transmission:}
Using the following Maclaurin series $e^x= 1 + x + o(x^2)$ and $(1-x)^{-1}=1+x +o(x^2)$, the asymptotic CDF of RV $\Psi _{sd}$ can be obtained as \eqref{SD-dir}.
\begin{equation}
{F_{{\Psi _{sd}}}}\left( {{\Psi _{sd}}} \right) =\sum\limits_{i = 1}^N {{\pi _{sd}}} \left( {1 + {{\bar \sigma }_{id}}} \right){\left( {{{\bar \sigma }_{sd}}} \right)^{ - 1}}{\Psi _{sd}} + o\left( {\bar \sigma _{sd}^{ - 1}} \right).\label{SD-dir}
\end{equation}
In turn, making use the above CDF, the following expression is obtained for the SOP in the asymptotic regime,
\begin{equation}
{\cal{P}}_{{\rm{out}}}^{\infty ,{\rm{DT}}} = \left[ {\sum\limits_{i = 1}^N {{\pi _{sd}}{2^{R}}{{\bar \sigma }_{se}}} \left( {1 + {{\bar \sigma }_{id}}} \right)} \right]{\left( {{{\bar \sigma }_{sd}}} \right)^{ - 1}}.\label{DT-high}
\end{equation}

\noindent \emph{DMC:}
We first note that in the asymptotic regime of high SNR, the probability of a WIRS event simplifies as follows,
\begin{align}\label{high-0}
\Pr \left( {{\cal F} = \emptyset } \right) = \prod\limits_{m = 1}^M {{{\cal{U}}_{\ell m }}} \bar \sigma _{sd}^{ - M}, \hspace{-4mm}
\quad {\rm{and}} \quad
 \hspace{-2mm}
\Pr \left( {{\cal F} = {\cal F}_n } \right)
 =  \prod\limits_{m \in {{\bar F}_n}}^{} {{{\cal{U}}_{\ell m }}}  \bar \sigma _{sd}^{\left| {{{\cal{F}}_n}} \right| - M},
\end{align}
\noindent where ${{{\cal{U}}_{\ell m }}}   = \sum\limits_{\ell  = 1}^N {{\pi _{s\ell m }}\hat \varepsilon } \left( {{{\bar \sigma }_{\ell m }} + 1} \right)\varrho $. Then, making use of \eqref{SD-dir} and \eqref{high-0} along with appropriate power series expansion, the asymptotic SOP of the DMC scheme is obtained as
\begin{align}
\hspace{-1mm}&{\cal{P}}_{{\rm{out}}}^{\infty ,{\rm{DMC}}}\!\! =
{\varepsilon ^{\left| {{{{\cal{F}}_n}}} \right|}}{2^{2R\left( {\left| {{{{\cal{F}}_n}}} \right| + 1} \right)}}\sum\limits_{i = 1}^N {\sum\limits_{m = 0}^{\left| {{{{\cal{F}}_n}}} \right| + 1} {\sum\limits_{l = 1}^2 {{\left| {{{{\cal{F}}_n}}} \right|+1} \choose m } } }\times\nonumber\\
&{\bf{g}}{\left( l \right)^{ - \left( {\left| {{{{\cal{F}}_n}}} \right|+ 2} \right)}}{\bf{h}}\left( l \right)\!{\pi _{sd}}\bar \sigma _{id}^m\Gamma\! \left( {\left| {{{{\cal{F}}_n}}} \right| + 1} \right)m!{\left( {{{\bar \sigma }_{sd}}} \right)^{ - \left( {\left| {{{{\cal{F}}_n}}} \right| + 1} \right)}}\!\!.\label{high-dmc-1}
\end{align}

By proceeding in a similar manner, we can obtain the SOP expressions of the other schemes, which are presented below




\noindent \emph{DSC:}
\begin{align}
&{\cal{P}}_{{\rm{out}}}^{\infty ,{\rm{DSC}}}\hspace{-1mm}=\hspace{-1mm}{\varepsilon ^{\left| {{{\cal{F}}_n}} \right|}}{2^{2R\left( {\left| {{{\cal{F}}_n}} \right| + 1} \right)}}\sum\limits_{m = 0}^{\left| {{{\cal{F}}_n}} \right| + 1} {\sum\limits_{i = 1}^N {\sum\limits_{l = 1}^3 \binom{{\left| {{{{\cal{F}}_n}}} \right|+1}}{m}{} } } {\bf{a}}\left( l \right)\times\nonumber\\
&\hspace{+3mm}{{{\bf{b}}}^{ - \left( {\left| {{{\cal{F}}_n}} \right| + 2} \right)}}\left( l \right){\pi _{sd}}m!\bar \sigma _{id}^m\Gamma \left( {\left| {{{\cal{F}}_n}} \right| + 2} \right){\left( {{{\bar \sigma }_{sd}}} \right)^{ - \left( {\left| {{{\cal{F}}_n}} \right| + 1} \right)}}.
\end{align}
%

\noindent \emph{DMM:}
\begin{equation}
{\cal{P}}_{{\rm{out}}}^{\infty ,{\rm{DMM}}}\!\!\!=\!\! \sum\limits_{i = 1}^N \!{\sum\limits_{m = 0}^2 \!{\sum\limits_{l = 1}^2\!\! {\binom{2}{m}\frac{\varepsilon{{\pi _{sd}}{\bf{\tilde h}}\bar \sigma _{id}^m}}{{2^{-4R}}{{{{\bf{\tilde g}}}^3}}}} \Gamma \!\left( {m\! +\! 1} \right)} } {\left( {{{\bar \sigma }_{sd}}} \right)^{ - 2}}\!\!.
\end{equation}


\noindent \emph{DSM:}
\begin{align}
\hspace{-2mm}{\cal{P}}_{{\rm{out}}}^{\infty ,{\rm{DSM}}}\hspace{-1mm}\!=\!
 \sum\limits_{i = 1}^N \!{\sum\limits_{m = 0}^2 \!{\sum\limits_{l = 1}^3 \!\!{\binom{2}{m}}}}\!\frac{\varepsilon {\bf{\tilde a}}\left( l \right)\!{\pi _{sd}}\bar \sigma _{id}^m\Gamma\! \left( {m \hspace{-1mm}+ \hspace{-1mm}1} \right)} {{2^{-4R - 1}}{{{\bf{\tilde b}}}^{3}}\left( l \right)} {\left( {{{\bar \sigma }_{sd}}} \right)^{ - 2}}\!\!.
\end{align}
%
%
%
\noindent \emph{DMA:}
Shown on the top of the next page.
\begin{figure*}
\small
\begin{align}
{\cal{P}}_{{\rm{out}}}^{\infty ,{\rm{DMA}}}=& {\varepsilon ^{\left| {{{\cal{F}}_n}} \right|}}\frac{{{2^{{\rm{2R}}\left( {\left| {{{\cal{F}}_n}} \right| + 1} \right)}}}}{{\Gamma \left( {\left| {{{\cal{F}}_n}} \right| + 2} \right)}}\sum\limits_{i = 1}^N {\sum\limits_{m = 0}^{\left| {{{\cal{F}}_n}} \right| + 1} {{\pi _{sd}}\bar \sigma _{id}^m\Gamma \left( {m + 1} \right)\binom{\left| {{{{\cal{F}}_n}}} \right|+1}{m}} }
 \times \nonumber\\&
 \left[ {q'\Gamma \left( {\left| {{{\cal{F}}_n}} \right| + 2} \right)\bar \sigma _{se}^{\left| {{{\cal{F}}_n}} \right| + 1} + \sum\limits_{k = 1}^{\left| {{{\cal{F}}_n}} \right|} {\frac{{{{\xi '}_k}\Gamma \left( {\left| {{{\cal{F}}_n}} \right| + k + 1} \right)}}{{\Gamma \left( k \right)}}} \bar \sigma _{me}^{\left| {{{\cal{F}}_n}} \right| + k + 1}} \right]{\left( {{{\bar \sigma }_{sd}}} \right)^{ - \left( {\left| {{{\cal{F}}_n}} \right| + 1} \right)}}.\label{dma-d}
\end{align}
\hrule
\end{figure*}

\noindent \emph{DSA:}
Shown on the top of the next page.
\begin{figure*}
\small
\begin{align}
{\cal{P}}_{{\rm{out}}}^{\infty ,{\rm{DSA}}}=&
{\varepsilon ^{\left| {{{\cal{F}}_n}} \right|}}{2^{{\rm{2R}}\left( {\left| {{{\cal{F}}_n}} \right| + 1} \right)}}\sum\limits_{i = 1}^N {\sum\limits_{m = 0}^{\left| {{{\cal{F}}_n}} \right| + 1} {\sum\limits_{{m_1} = 1}^{\left| {{{\cal{F}}_n}} \right|} {{{\left( { - 1} \right)}^{{m_1} - 1}}\frac{{{m_1}{\pi _{sd}}\bar \sigma _{id}^m\Gamma \left( {m + 1} \right)\Gamma \left( {\left| {{{\cal{F}}_n}} \right| + 2} \right)\xi }}{{{{\bar \sigma }_{me}}{{\bar \sigma }_{se}}}}} } }  \times
\binom{\left| {{{{\cal{F}}_n}}} \right|}{m_1} \binom{\left| {{{{\cal{F}}_n}}} \right|+1}{m}\nonumber\\&
\times \left[ {{{\left( {\frac{{{{\bar \sigma }_{me}}}}{{{m_1}}}} \right)}^{\left| {{{\cal{F}}_n}} \right| + 2}} - \bar \sigma _{se}^{\left| {{{\cal{F}}_n}} \right| + 2}} \right]{\left( {{{\bar \sigma }_{sd}}} \right)^{ - \left( {\left| {{{\cal{F}}_n}} \right| + 1} \right)}}\label{dsa-d}.
\end{align}
\hrule
\end{figure*}

\noindent \emph{DMO:}
\begin{align}
\hspace{-2mm}&{\cal{P}}_{{\rm{out}}}^{\infty ,{\rm{DMO}}} = {\left( {\frac{\varepsilon }{{\tilde \varepsilon }}} \right)^{\left| {{{\cal{F}}_n}} \right|}}\bar \sigma _{se}^{\left| {{{\cal{F}}_n}} \right| + 1}\frac{{{2^{2R\left( {\left| {{{\cal{F}}_n}} \right| + 2} \right)}}}}{{\left( {\left| {{{\cal{F}}_n}} \right| + 1} \right)\left( {\left| {{{\cal{F}}_n}} \right| + 2} \right)}}\times\nonumber\\
&\sum\limits_{m = 0}^{\left| {{{\cal{F}}_n}} \right|}\! {\sum\limits_{i = 1}^N \!\!{{\pi _{sd}}\bar \sigma _{id}^m} \binom{\left| {{{{\cal{F}}_n}}} \right|}{m}} \!\left( {m! + {{\bar \sigma }_{id}}\left( {m + 1} \right)!} \right)\bar \sigma _{sd}^{ - \left( {\left| {{{\cal{F}}_n}} \right| + 1} \right)}.
\end{align}


\noindent \emph{DSO:}

\begin{align}
{\cal{P}}_{{\rm{out}}}^{\infty ,{\rm{DSO}}} \hspace{-1mm}=& {\left( {\frac{\varepsilon }{{\tilde \varepsilon }}} \right)^{\left| {{{\cal{F}}_n}} \right|}}\bar \sigma _{se}^{\left| {{{\cal{F}}_n}} \right| + 1}{2^{2R\left( {\left| {{{\cal{F}}_n}} \right| + 2} \right)}}\sum\limits_{m = 0}^{\left| {{{\cal{F}}_n}} \right|} {\sum\limits_{i = 1}^N {{\pi _{sd}}\bar \sigma _{id}^{m + 1}}}\times\nonumber\\
&
{\binom{\left| {{{{\cal{F}}_n}}} \right|}{m}}
 \left( {m! + \bar \sigma _{id}^{}\left( {m + 1} \right)!} \right)\bar \sigma _{sd}^{ - \left( {\left| {{{\cal{F}}_n}} \right| + 1} \right)}.
\end{align}
%
%


\begin{table}
\begin{center}
\caption{Diversity order in high SNR regime}
\vspace{-1mm}
\begin{tabular}{| c| c| c| c| c| c| c| c| c| c}
\hline
Scheme &  Diversity order \\
\hline
  DT & $1$ \\
\hline
  DMC  & $M+1$ \\
  \hline
  DSC & $M+1$   \\
 \hline
  DMM  & $M-\left| {{{\cal{F}}_n}} \right|+2$  \\
  \hline
  DSM   & $M-\left| {{{\cal{F}}_n}} \right|+2$  \\
  \hline
  DMO  &  $M+1$   \\
  \hline
  DSO &  $M+1$   \\
  \hline
  DMA & $M+1$  \\
  \hline
  DSA & $M+1$    \\
  \hline
\end{tabular}
\vskip -15pt
\label{table:dv}
\end{center}
\end{table}


\subsection{Diversity Order and Coding Gain}
In the high SNR regime for the legitimate links, the coding gain and diversity order are defined through the obtained asymptotic expression for the SOP, that is: ${P_{{\rm{out}}}^{\infty ,Q}}\approx \left(C{\bar \sigma }_{sd}\right)^{-D}$, where $C$ and $D$ respectively denote the coding gain and  the diversity order of the scheme $Q$ under consideration. For example, in the case of $DT$, we immediately obtain from \eqref{DT-high} that
\begin{equation}
C = \left[ {\sum\limits_{i = 1}^N {{\pi _{sd}}{2^{R}}{{\bar \sigma }_{se}}} \left( {1 + {{\bar \sigma }_{id}}} \right)} \right]^{-1}, ~D = 1.
\end{equation}
Proceeding in this manner, we can obtain the coding gain and diversity order for each one of the schemes considered in Subsection A.
For reference, the diversity orders of these schemes are listed in Table II, while the coding gain can easily be computed as
\begin{equation}
C_Q=\frac{{\left({\cal{P}}_{{\rm{out}}}^{\infty ,{\rm{Q}}}\right)}^{-\frac{1}{D_Q}}}{{{{\bar \sigma }_{sd}}} },
\end{equation}
with the corresponding expression for ${\cal{P}}_{{\rm{out}}}^{\infty ,{\rm{Q}}}$ calculated previously. Based on the above
results, the diversity order of the schemes are summarized in Table~\ref{table:dv}.

\subsection{Remarks}

\begin{itemize}
  \item The maximum diversity order of $M+1$ is achieved for the DMC, DSC, DMA, DSA, DMO and DSO schemes. In contrast, the DMM and DSM scheme achieve a (conditional) diversity order of $M-\left| {{{\cal{F}}_n}} \right|+2$, which decreases with the cardinality of the WIRS. Finally, the worst performing scheme is DT with a diversity order of $=1$.

  \item Since the diversity orders of DMC, DSC, DMA, DSA, DMO, and DSO are identical, the tradeoff among these schemes is solely characterized by their respective coding gains. Hence, their relative performance can be quantified in terms of the simple ratio of their coding gains, which can be interpreted as an SNR gap. For example, the SNR gap between the DMC and DSC schemes is given by
      \begin{align}
      \Delta C=\frac{C_{{\rm{DMC}}}}{C_{{\rm{DSC}}}}.
      \end{align}
Here $C_{{\rm{DMC}}} > C_{{\rm{DSC}}} $ and so $\Delta C >1$, indicating that DMC outperforms DSC by $20\log_{10}\Delta_{{\rm{C}}}$ for the same SOP. We can show that the relative performance of the above schemes can be ordered as $C_{{\rm{DMO}}}>C_{{\rm{DSO}}}>C_{{\rm{DMC}}}>C_{{\rm{DSC}}}>C_{{\rm{DMA}}}>C_{{\rm{DSA}}}$.
  \item It is observed that when $\mathcal{F}_n=\varnothing$, \emph{i.e.}, no relay can decode the source symbol successfully, the exact SOP for all scenarios is reduced to
  \begin{align}
  {\cal{P}}_{{\rm{out}}} = \left[ {\sum\limits_{i = 1}^N {{\pi _{sd}}{2^{R}}{{\bar \sigma }_{se}}} \left( {1 + {{\bar \sigma }_{id}}} \right)} \right]{\left( {{{\bar \sigma }_{sd}}} \right)^{ - 1}}.
  \end{align}

In the special case when both $\sigma_{sd}$ and $\sigma_{se}$ $\rightarrow \infty$ at the same rate, the above expression results in a constant SOP; in turn, this floor phenomenon leads to a zero diversity gain.
\end{itemize}

\section{Numerical Results and Discussions}
In this section, we present numerical results to validate the derived theoretical expressions of the SOP for the proposed methods. In the simulations, the noise variances are all normalized to unity, the data rate $\cal{R}$ = $0.5$ bits per channel use. The simulation results are obtained by averaging over $10^5$ independent runs, and the number of transmitted bits is set to $10^4$ for each run. The relative SNRs of the various links are characterized by the following parameter values: $\varepsilon  = 1.01$, ${{\tilde \varepsilon }} = 0.9$ and $\hat \varepsilon  = 1.03$. The default values of the parameters $M$ and $N$, i.e. number of relays and eavesdropper antennas, are set to $4$ and $5$, respectively, unless otherwise specified.

\begin{figure}
\centering
\includegraphics[width=3.5 in]{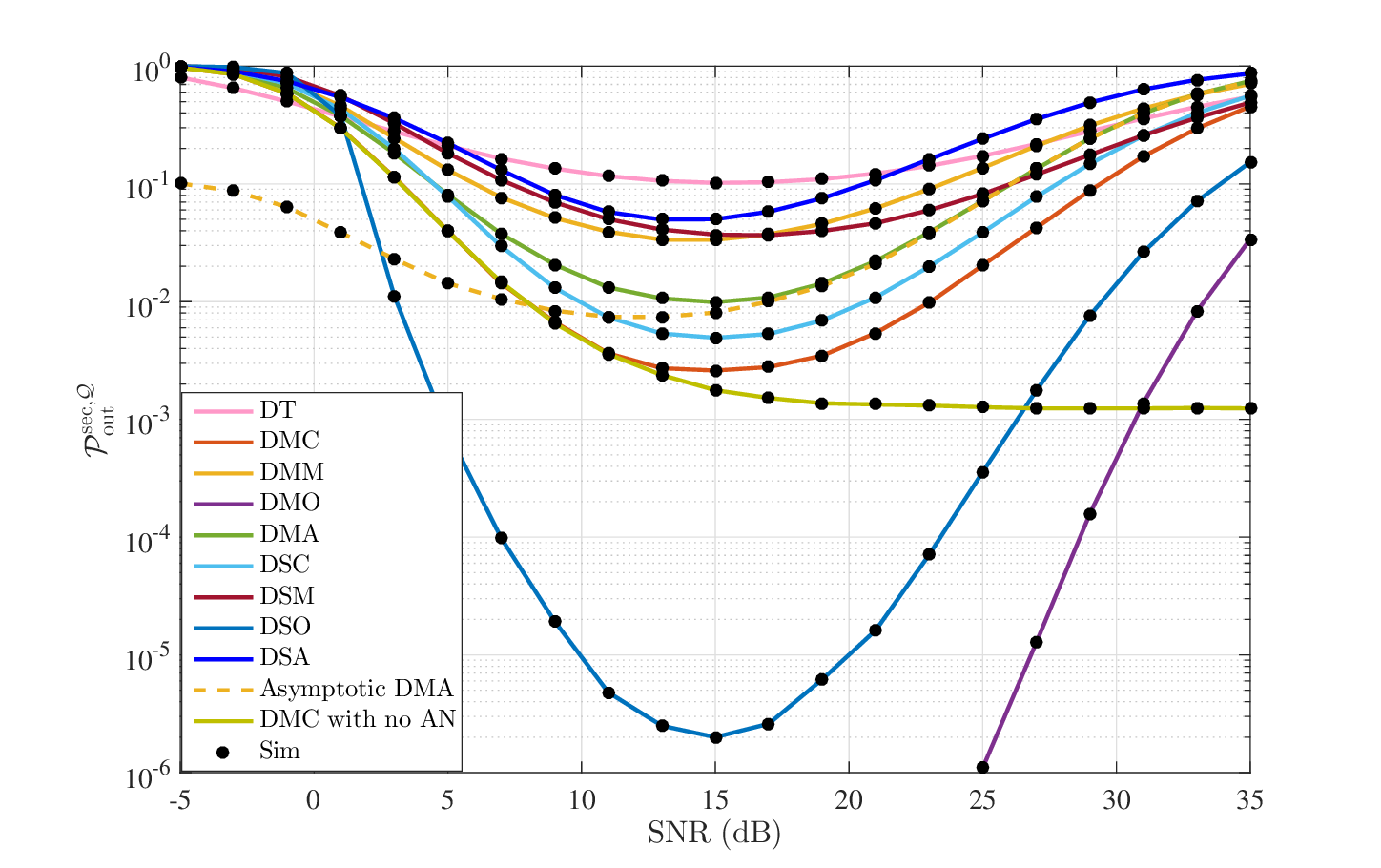}
\caption{The secrecy outage performance of proposed relay selection schemes and direct transmission versus SNR ( $M=4$ and $N=5$).}
\label{SOP_SNR}
\end{figure}

Fig.~\ref{SOP_SNR}, shows the SOP comparison among the DT, DMC, DSM, DMM, DMA, DSO, DMO, DSC and DSA, by plotting Eqs. \eqref{Pr_SD_DF}, \eqref{outage probability-DMC}, \eqref{outage probability-DSC}, \eqref{outage probability-DMM}, \eqref{outage probability-DSM}, \eqref{C-sd-DSo}, \eqref{DMO-op}, \eqref{outage probability-DSA} and \eqref{outage probability-DMA}, respectively, as  function of SNR. From the various curves in Fig.~\ref{SOP_SNR}, it is seen that the DMC, DSM, DMM, DMA, DSO, DMO, DSC and DSA schemes (in the low to medium SNR range) all perform better than DT in terms of secrecy performance, demonstrating the security benefits of exploiting cooperative relays to defend against eavesdropping attack. One can also see from Fig.~\ref{SOP_SNR} that the SOP performance of the DMO and DSO schemes is better than that of the other schemes, thereby showing the advantage of the optimal relay selection over the other selection schemes and multiple relay selection, as well as the traditional DT. The figure also includes a special curve for the case where the number of transmit antennas of the eavesdropper node $N=0$ corresponding to the case where $E$ does not send any AN to the legitimate network. From the simulation results, we see that the error floor phenomenon occurs in the absence of AN. This observation confirms that the presence of AN can highly affect the secrecy outage performance of the legitimate network. As can be observed, the analytical results are in perfect agreement with the simulation results, which demonstrates the validity of the derived analytical expressions. We also find that the high SNR approximations are quite accurate (although the asymptotic result is only plotted for DMA in order not to cause confusion in the figure).


%
%
%

\begin{figure}
\centering
\includegraphics[width=3.5 in]{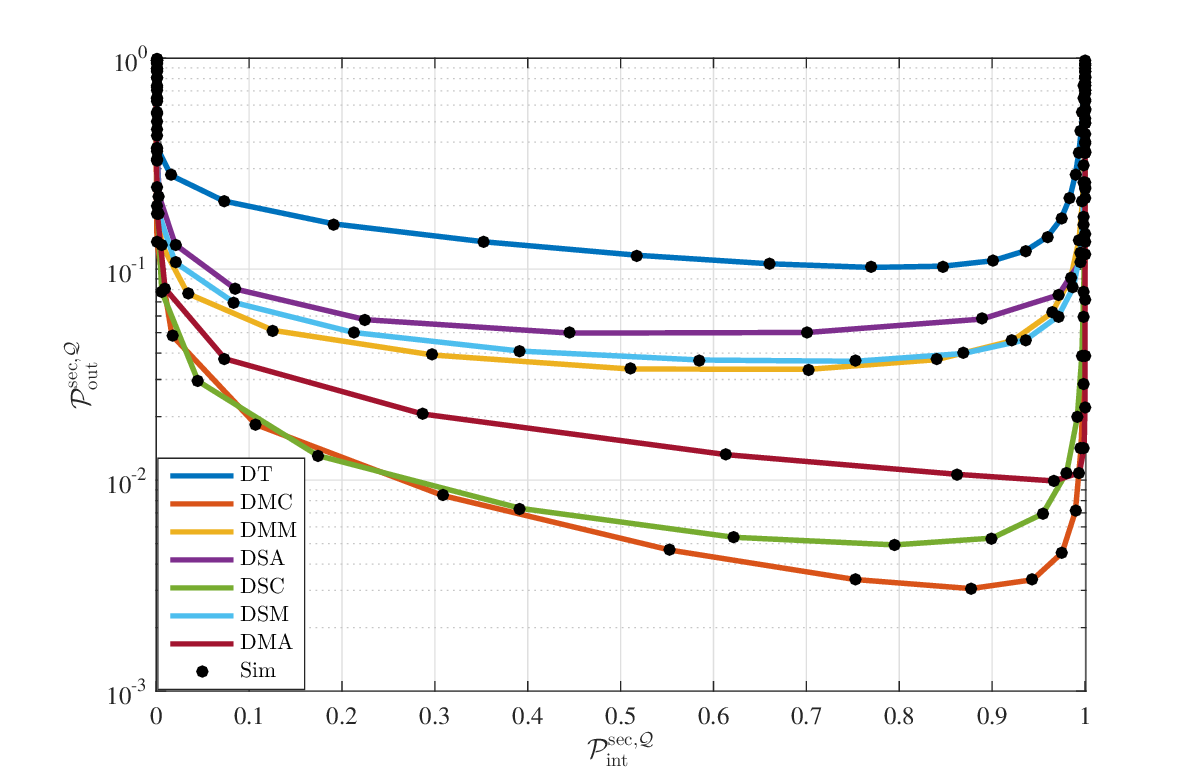}
\caption{The secrecy outage performance of proposed relay selection schemes and direct transmission as a function of intercept probability ( $M=4$ and $N=5$).}
\label{SOP_IP}
\end{figure}


\begin{figure}
\centering
\includegraphics[width=3.5 in]{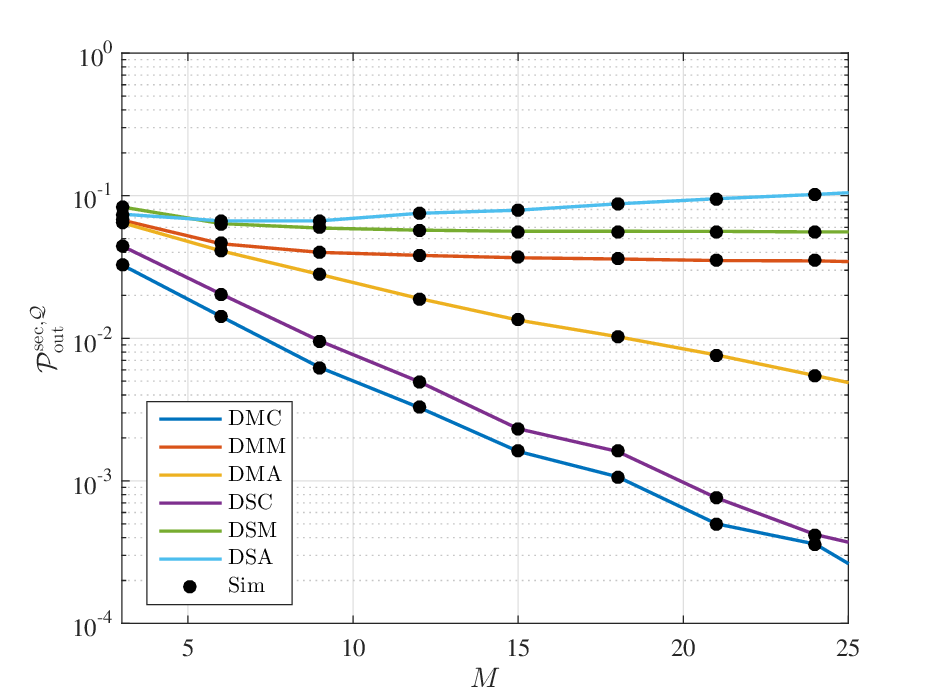}
\caption{The secrecy outage performance of proposed relay selection schemes and direct transmission as a function of the number of relays $M$ ($\rm{SNR} = 10dB, N=5$).}
\label{SOP_M}
\end{figure}


\begin{figure}
\centering
\includegraphics[width=3.5 in]{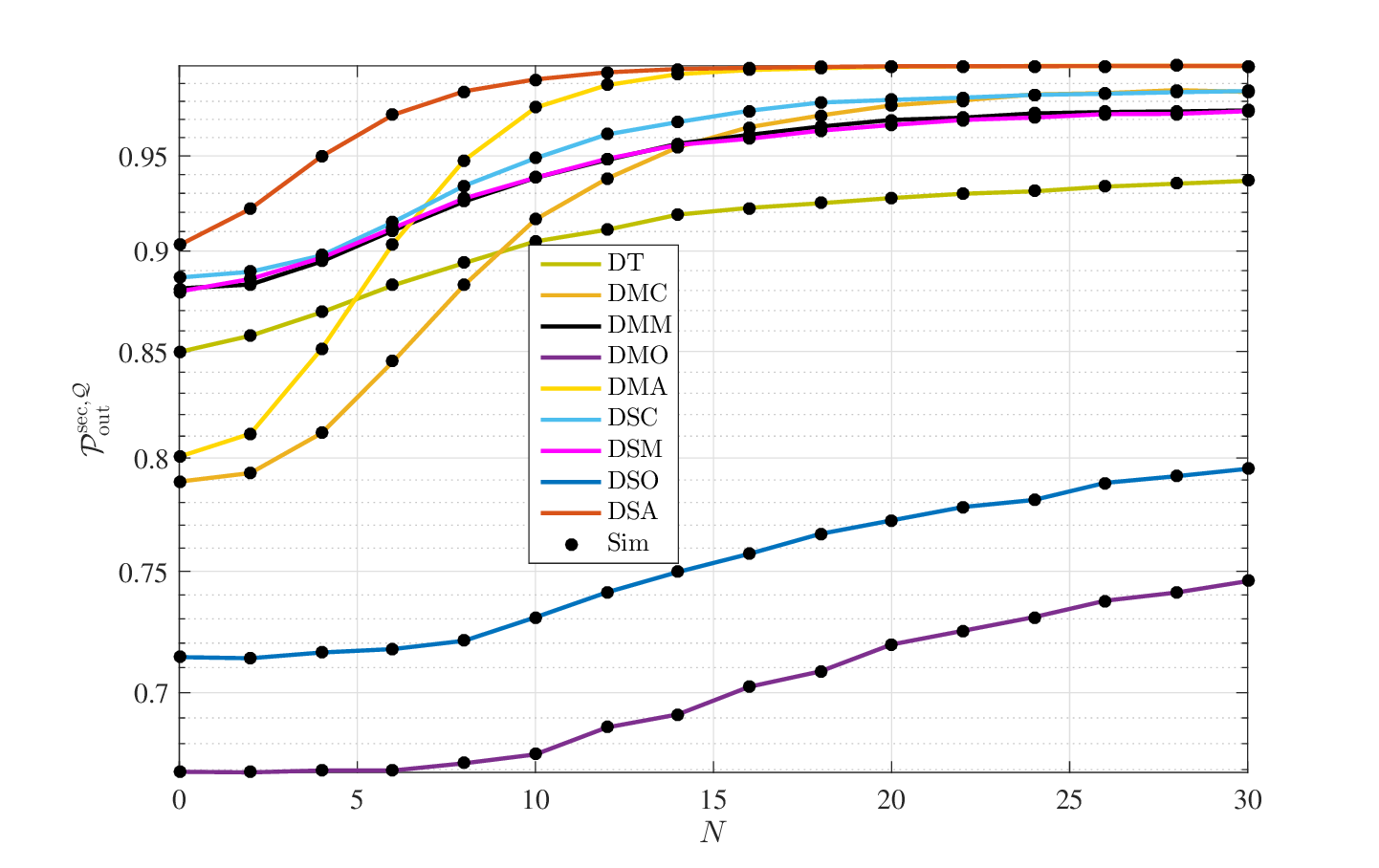}
\caption{The secrecy outage performance of proposed relay selection versus the the number of eavesdropper $N$ ($M$=4, \rm{SNR}=30dB).}
\label{SOP_N}
\end{figure}

Fig.~\ref{SOP_IP} shows numerical SOP results versus IP for both the conventional DT and the proposed single and multi-relay selection schemes, where the legitimate-to-eavesdropper channel gain ratio is around 11 dB. One can see from Fig.~\ref{SOP_IP} that for a specific IP value, the SOP of the proposed relay selection schemes is strictly lower than that of DT, thereby confirming that the former outperform the conventional DT. It can be observed that the DSC and DMC schemes outperform the DSA approach ( i.e., when all successful relays in the WIRS are involved in transmission). This can be explained by noting that in the DSA case, the eavesdropper receives multiple copies of the source signal when multiple relays transmit, which in turn degrades the secrecy performance.


Fig.~\ref{SOP_M} shows the SOP as a function of the number for relays $M$ of the DMC, DSM, DSA, DMM, DMA, and DSC schemes. It is observed from Fig.~\ref{SOP_M} that the DMC scheme performs better than the other single and multi-relay selection schemes in terms of SOP, except the DSO and DMO. Again the proposed optimal relay selection schemes, DSO and DMO, outperform the other schemes. Since even with a small increase in $M$ the SOP of DMO and DSO rapidly tends to zero, the corresponding curves are not sketched here. One can also see from Fig.~\ref{SOP_M} that as the number of relays $M$ increases, the SOP of the various schemes rapidly decreases, except for the DSA, DSM and DMM schemes. In the case of DSA, this can be explained by noting that for a fair comparison, the total amount of transmit power at the source and relay shall be limited to $P^{'}=2P_m=2P_s$. However, using the equal-power allocation for simplicity, the transmit power at the source and relay is given by $P_m=P_s$. Thus, in DSA where all the relays in the WIRS cooperate in the next phase, the power of each relay is reduced to $P^{'}/\left( {\left|{{\cal{F}}_n}\right|+1} \right)$ which negatively affects the secrecy performance. The same line of thoughts can be applied for DMA; however, in this case with an increasing number of relays, the secrecy performance improves. In addition, as shown in Fig.~\ref{SOP_M}, the SOP improvement of DSC and DMC becomes more significant as $M$ increases.


Fig.~\ref{SOP_N} examines the impact of the number $N$ of eavesdropper's transmit antennas on the secrecy outage performance when $M=4$, $\rm{SNR} = 30 dB$ and the legitimate-to-eavesdropper channel gain ratio is around $-3$ dB. The results clearly demonstrate that wireless security degrade with an increase in the number $N$ of antennas. Furthermore, for both MRC and SC combining solutions, the proposed optimal relay selection outperforms the other relay selection schemes as well as the DT in terms of SOP. In other words, the DMO and DSO schemes
achieves the best SOP performance, further confirming the advantage of the proposed optimal relay selection. That is, no matter which combining solution (\emph{i.e.}, MRC and SC) is considered, the proposed optimal relay selection always performs better than the traditional relay selection and multiple relay combining approaches in terms of secrecy performance. We see that at the low SNRs, both MRC and SC combining solutions achieve similar results, while at the high SNRs, the MRC scheme outperforms the SC scheme.

%
%





\section{Concluding Remarks}

We have studied the PHY-layer security in a cooperative wireless subnetwork that includes a source-destination pair, multiple relays, and a malevolent active eavesdropper, which can transmits AN with a multiple-antenna transmitter to degrade the achievable secrecy rate of the legitimate channels. Depending on the availability of the CSI, we considered different relay selection schemes, i.e. conventional, minimum and optimal selection, along with different combining methods at the destination and eavesdropper, i.e. MRC and SC. We first analyzed the secrecy capacity of the direct transmission in terms of IP and SOP. A DF incremental relaying (IR) protocol was then introduced to improve reliability and security of communications in the presence of the eavesdropper. For each one of the proposed relay selection schemes, and for both MRC and SC, we derived new and exact closed-form expressions for the IP and SOP under the Rayleigh channel assumption. We also characterized the secrecy performance of the various relay schemes in the asymptotic high SNR regime, which enabled us to obtain the associated coding gains and diversity orders. For both signal combining solutions, the proposed IR schemes (except DMM and DSC) achieve the maximum diversity order of $M+1$, where $M$ is the number of relays. Our analysis and simulation results have revealed that the IR-based relaying with optimal selection outperforms the conventional selection, which in turns outperforms minimum selection. Our results also indicate that as $M$ increases, the secrecy performance of the DMO, DSO, DMC, DSC and DMA schemes improves rapidly.



\appendices
\section{Proof of Lemma~\ref{CDF-Y}}
\label{App-CDF-Y}
Let us introduce the following intermediate random variables (RV): $Y={\mathop {\max }\limits_{m\, \in \,{{{{\cal{F}}_n}}}} {\Psi _{md}} + {\Psi _{sd}}}$, ${\gamma _{id}} = \frac{{{P_{id}}{{\left| {{c_{i,d}}} \right|}^2}}}{{\sigma _n^2}}$, ${\gamma _D} = \sum\limits_{i = 1}^N {{\gamma _{id}}} $ and $X = \mathop {\max }\limits_{m\, \in \,{{{{\cal{F}}_n}}}} {\Psi _{md}}$. The existence of the common RV ${{\gamma _D}}$ in $X$ and $Y$ leads to a statistical dependence between RVs $X$ and $Y$. By conditioning on ${{\gamma _D}}$, we first obtain \vspace{-4mm}

\begin{align}
{F_Y}(\gamma ) &= {E_{{\gamma _D}}}\left[ {\Pr \left( {X + V \le \gamma \left( {{\gamma _D} + 1} \right)\left| {{\gamma _D}} \right.} \right)} \right]\\&={E_{{\gamma _D}}}\left[ {{E_V}\left[ {{F_X}\left( {\gamma \left( {{\gamma _D} + 1} \right) - V} \right)\left| V \right.} \right]\left| {{\gamma _D}} \right.} \right].
\label{F-Yy}
\end{align}

\noindent Using of the binomial theorem, we obtain the CDF of $X$ and the PDF of ${{\gamma _D}}$ as
\begin{align}
{F_X}(\gamma ) &= 1 - \sum\limits_{m = 1}^{\left| {{{{\cal{F}}_n}}} \right|} {{{( - 1)}^{m - 1}}\binom{\left| {{{{\cal{F}}_n}}} \right|}{m}\exp ( - \frac{{m\gamma }}{{{{\bar \sigma }_{md}}}})},\quad {\rm{}} \quad\nonumber\\
{f_{{\gamma _D}}}(\gamma ) &= \,\sum\limits_{i = 1}^N {\frac{{{\pi _{sd}}}}{{{{\bar \sigma }_{id}}}}} \exp \left( { - \frac{\gamma }{{{{\bar \sigma }_{id}}}}} \right).
\label{PDF-gammaD}
\end{align}

\noindent Then, with the help of \eqref{CDF-X}, \eqref{F-Yy} and \eqref{PDF-gammaD}, and using properties of conditional expectations~\citep{Papoulis:1984}, we finally arrive at the expression of ${F_Y}(\gamma ) $ in \eqref{F_Y}.

\section{Proof of Theorem~\ref{Thm-outage probability-DMC}}
\label{App-DMC-proof}
Introducing $Z={{\Psi _{{m }e}} + {\Psi _{se}}}$ and according to the definition of SOP, we have
\begin{align}
{\cal{P}}_{{\rm{out}}}^{{\rm{DMC}}} &=\hspace{-1mm} \Pr \left( {Y < \varrho  \hspace{-1mm}+\hspace{-1mm} \left( {\varrho  \hspace{-1mm}+\hspace{-1mm} 1} \right)Z} \right) = {{\rm{E}}_Z}\left[ {{F_Y}\left( {\varrho  + \left( {\varrho  + 1} \right)Z} \right)} \right]\nonumber \\&
 = \Pr \left( {C_{sd}^{{\rm{DMC}}} < {\cal{R}}\left| {Y > Z} \right.} \right)\Pr \left( {Y > Z} \right)\nonumber \\&
  + \Pr \left( {C_{sd}^{{\rm{DMC}}} < {\cal{R}}\left| {Y < Z} \right.} \right)\Pr \left( {Y < Z} \right). \label{App-DMC}
\end{align}
It is straightforward to verify that $\Pr \left( {C_{sd}^{{\rm{DMC}}} < {\cal{R}}\left| {Y < Z} \right.} \right)=1$. Then, the first term in \eqref{App-DMC} can be expressed as
\begin{align}
&\Pr \!\left( {C_{sd}^{{\rm{DMC}}}\! <\! {\cal{R}}\left| {Y\! > \!Z}\! \right.} \right)\Pr \left(\! {Y\! >\! Z} \right) \!=
 \!\frac{{\Pr \left( {C_{sd}^{{\rm{DMC}}} < {\cal{R}},Y\! >\! Z} \right)}}{{\Pr \left( {Y \!>\! Z} \right)}}\times\nonumber\\
 &\hspace{10mm}\Pr \left( {Y > Z} \right) =\Pr \left( {Z < Y < {2^{2{\cal{R}}}}Z + \varrho} \right)\nonumber\\
 &\hspace{10mm} =  \Pr \left( {Y < {2^{2{\cal{R}}}}Z + \varrho} \right) - \Pr \left( {Z < Y} \right). \label{DMC-E}
\end{align}
Making use of \eqref{DMC-E} and \eqref{App-DMC} we can write
\begin{align}
\hspace{-2mm}{\cal{P}}_{\rm{out}}^{{\rm{DMC}}}  \hspace{-1mm}=\hspace{-1mm} \Pr \left( {Y \!<\! \varrho  +\hspace{-1mm} \left( {\varrho\!  +\! 1} \right)\!Z} \right)\! = \!{{\rm{E}}_Z}\!\left[ {{F_Y}\!\left( {\varrho  +\hspace{-1mm} \left( {\varrho\!  + \!1} \right)Z} \right)} \right].
\label{C-sd-DMC}
\end{align}
To prove the desired result, we call upon \eqref{C-sd-DMC} and exploit the pdf of the RV $Z$ as
\begin{equation}
{f_Z}(\gamma ) = \sum\limits_{l = 1}^2 {{\bf{h}}\left( l \right)\exp \left( { - {\bf{g}}\left( l \right)\gamma } \right)}.
\label{fzz}
\end{equation}
Then, according to \eqref{C-sd-DMC}, \eqref{fzz} and conjuring the identity \citep[Eq. (2.1.3.1)]{Wolfram:website} we arrive at ${\cal{P}}_{\rm{out}}^{{\rm{DMC}}}$ as in \eqref{outage probability-DMC}.

\section{Proof of Theorem~\ref{Thm-outage probability-DMM}}
\label{App-DMMM-proof}
The desired SOP can be first expressed in terms of the CDF of RV $\tilde Y $
\begin{equation}
\hspace{-2mm}{\cal{P}}_{\rm{out}}^{{\rm{DMM}}} \hspace{-1mm}=\hspace{-1mm} \Pr\! \left( \!{\tilde Y \!<\! \varrho \! +\! \left( {\varrho  \hspace{-1mm} +\hspace{-1mm} 1} \right)\!\tilde Z} \!\right) \hspace{-1mm}= \hspace{-1mm}{{\rm{E}}_{\tilde Z}}\left[ {{F_{\tilde Y}}\left( {\varrho  \hspace{-1mm}+\hspace{-1mm} \left( {\varrho  \hspace{-1mm}+\hspace{-1mm} 1} \right)\tilde Z} \right)} \right].
\label{CDF-YZ}
\end{equation}

\noindent where $\tilde Z= {\mathop {\min }\limits_{m\, \in \,{{\cal{F}}_n}} {\Psi _{me}}+{\Psi _{se}}}$ with its PDF given by
\begin{equation}
{f_{\tilde Z}}(z) = \sum\limits_{l = 1}^2 {{\bf{\tilde h}}\left( l \right)\exp \left( { - {\bf{\tilde g}}\left( l \right)z} \right)}.
\label{vvvc}
\end{equation}
Then, making use of \eqref{F_Y_tilda}, \eqref{CDF-YZ}, and \eqref{vvvc} along with the identity \citep[Eq. (2.1.3.1)]{Wolfram:website}, we finally obtain \eqref{outage probability-DMM}.




\section{Proof of Lemma~\ref{Lem-CDF-Psi-md-DMA}}
\label{App-lem-DMM}
Introducing the $\Delta  = \sum\limits_{m \in {{{{\cal{F}}_n}}}}^{} {{\gamma _{md}}} ,{\gamma _{md}} = {P_m}{\left| {{h_{md}}} \right|^2}$, ${\gamma _d} = \sum\limits_{i = 1}^{{N}} {{\gamma _{id}}} $ and ${\gamma _{id}} = \frac{{{P_{id}}{{\left| {{c_{id}}} \right|}^2}}}{{\sigma _n^2}}$ we have
\begin{equation}
{F_{\Psi _{md}^{{\rm{DMA}}}}}\left( \varrho  \right) = {{\rm{E}}_{{\gamma _d}}}\left[ {\Pr \left( {\Delta  < \varrho {\gamma _d} + \varrho \left| {{\gamma _d}} \right.} \right)} \right].
\end{equation}
Making use of the moment generating function (MGF) of RVs $\Delta$, ${F_\Delta }\left( x \right)$ and ${f_{{\gamma _d}}}({\gamma _d})$ can be obtained as
\begin{align}
&{F_\Delta }\left( x \right) = 1 - \sum\limits_{k = 0}^{\left| {{{{{\cal{F}}_n}}}} \right| - 1} {{{\left( {\frac{x}{{{{\bar \sigma }_{md}}}}} \right)}^k}\frac{{\exp \left( { - \frac{1}{{{{\bar \sigma }_{md}}}}x} \right)}}{{k!}}}
,\quad {\rm{}} \quad\nonumber\\
&
{f_{{\gamma _d}}}({\gamma _d}) = \sum\limits_{i = 1}^{{N}} {\frac{{{\pi _{sd}}}}{{{{\bar \sigma }_{id}}}}} \exp \left( { - \frac{{{\gamma _d}}}{{{{\bar \sigma }_{id}}}}} \right).
\label{f-gamma-d}
\end{align}
Then, according to \eqref{f-gamma-d} we have
\begin{align}
&{F_{\Psi _{md}^{{\rm{DMA}}}}}\left( \varrho  \right)=
{{\rm{E}}_{{\gamma _d}}}\left[ {{F_\Delta }\left( {\varrho {\gamma _d} + \varrho } \right)} \right] = 1 - \int_0^\infty  {} \sum\limits_{k = 0}^{\left| {{{{{\cal{F}}_n}}}} \right| - 1} {\sum\limits_{i = 1}^N {\frac{{{\pi _{sd}}}}{{{{\bar \sigma }_{id}}}}} }\times\nonumber\\
& {\mkern 1mu}
\frac{{{{\left( {{\gamma _d} + 1} \right)}^k}{{\left( {\frac{\varrho }{{{{\bar \sigma }_{md}}}}} \right)}^k}\!\!\exp \left( { - {\gamma _d}\left( {\frac{\varrho }{{{{\bar \sigma }_{md}}}}\! + \frac{1}{{{{\bar \sigma }_{id}}}}} \right) - \frac{\varrho }{{{{\bar \sigma }_{md}}}}} \right)}}{{k!}}d{\gamma _d}.
\end{align}
Finally, the desired result is obtained by evaluating the above integral.
\section{Proof of Theorem~\ref{Thm-outage probability-DMA}}
\label{Apx-Thm-outage probability-DMA}
We first use the MGF to compute the PDF of the RV ${\Psi _{me}^{{\rm{DMA}}}}$ as
\begin{equation}
{f_{\Psi _{me}^{{\rm{DMA}}}}}  \left( \gamma  \right)\hspace{-1mm}=\hspace{-1mm} q'\! \exp\! \left(\! { -\! \frac{\gamma}{{{{\bar \sigma }_{se}}}} } \!\right)\! +\!\! \sum\limits_{q = 1}^{\left| {{{\cal{F}}_n}} \right|} \!{\frac{{{\xi' _q}}{\gamma ^{q - 1}}}{{\Gamma \left( q \right)}}} \exp \!\left( \!{ - \!\frac{\gamma}{{{{\bar \sigma }_{me}}}} } \!\right)\!.
\label{pdf-psi-me-DMA}
\end{equation} \noindent Then, based on \eqref{CDF-Psi-md-DMA}, \eqref{OP-DMA-D} and \eqref{pdf-psi-me-DMA}, we can obtain the following integral expression,
\begin{align}
\Pr \left( {C_{sd}^{{\rm{DMA}}} \hspace{-1mm}<\hspace{-1mm} R} \right) \hspace{-1mm}=\!\hspace{-1mm} \int_0^\infty\!\!\!\!  {{F_{\Psi _{md}^{{\rm{DMA}}}}}\!\left( {\varrho  \hspace{-1mm}+\hspace{-1mm} \left( {\varrho \! + \!1} \right)\!\gamma } \right)\!{f_{\Psi _{me}^{{\rm{DMA}}}}}\!\left( \gamma  \right)d\gamma }.
\end{align}
Finally, by invoking the binomial theorem as well as the identity \citep[Eq. (2.1.3.1)]{Wolfram:website}, we arrive at \eqref{outage probability-DMA}.










\bibliographystyle{IEEEtran}
\bibliography{IEEEabrv,refs_Mohammad}

\begin{thebibliography}{10}
\providecommand{\url}[1]{#1}
\csname url@samestyle\endcsname
\providecommand{\newblock}{\relax}
\providecommand{\bibinfo}[2]{#2}
\providecommand{\BIBentrySTDinterwordspacing}{\spaceskip=0pt\relax}
\providecommand{\BIBentryALTinterwordstretchfactor}{4}
\providecommand{\BIBentryALTinterwordspacing}{\spaceskip=\fontdimen2\font plus
\BIBentryALTinterwordstretchfactor\fontdimen3\font minus
  \fontdimen4\font\relax}
\providecommand{\BIBforeignlanguage}[2]{{%
\expandafter\ifx\csname l@#1\endcsname\relax
\typeout{** WARNING: IEEEtran.bst: No hyphenation pattern has been}%
\typeout{** loaded for the language `#1'. Using the pattern for}%
\typeout{** the default language instead.}%
\else
\language=\csname l@#1\endcsname
\fi
#2}}
\providecommand{\BIBdecl}{\relax}
\BIBdecl

\bibitem{Shannon:1949}
C.~Shannon, ``Communication theory of secrecy systems,'' \emph{Bell Syst.
  Technical J.}, vol.~28, no.~4, pp. 656--715, 1949.

\bibitem{Wyner:1975}
A.~D. Wyner, ``The wire-tap channel,'' \emph{Bell Sys. Tech. J}, vol.~54,
  no.~8, pp. 1355--1387, 1975.

\bibitem{Csiszar:Trans:1978}
I.~Csisz{\'a}r and J.~K{\"o}rner, ``Broadcast channels with confidential
  messages,'' \emph{IEEE Trans. Inform. Theory}, vol.~24, no.~3, pp. 339--348,
  1978.

\bibitem{ElGamal:TIF:2012}
O.~O. Koyluoglu, C.~E. Koksal, and H.~E. Gamal, ``On secrecy capacity scaling
  in wireless networks,'' \emph{IEEE Trans. Inf. Theory}, vol.~58, no.~5, pp.
  3000--3015, 2012.

\bibitem{gamal1}
K.~Khalil, O.~O. Koyluoglu, H.~E. Gamal, and M.~Youssef, ``Opportunistic
  secrecy with a strict delay constraint,'' \emph{IEEE Trans. on Commun.},
  vol.~61, no.~11, pp. 4700--4709, November 2013.

\bibitem{SaeedH2}
S.~Hajizadeh and G.~A. Hodtani, ``Three-receiver broadcast channels with side
  information,'' in \emph{Proceedings of {IEEE} International Symposium on
  Information Theory, {ISIT}, Cambridge, MA, USA, July}, 2012, pp. 393--397.

\bibitem{SaeedH1}
S.~Hajizadeh and N.~Devroye, ``Dependence balance outer bounds for the discrete
  memoryless two-way multiple access broadcast channel,'' in \emph{52nd Annual
  Allerton Conference on Communication, Control, and Computing, Allerton, IL,
  USA}, 2014, pp. 996--1003.

\bibitem{gamal2}
L.~Lai and H.~E. Gamal, ``The relay-eavesdropper channel: Cooperation for
  secrecy,'' \emph{IEEE Transactions on Information Theory}, vol.~54, no.~9,
  pp. 4005--4019, Sept 2008.

\bibitem{Cheong:Thesis:1976}
S.~K. Leung-Yan-Cheong, ``Multi-user and wiretap channels including feedback,''
  \emph{Ph.D. thesis, Stanford University, Stanford, CA}, 1976.

\bibitem{Soysal:TVT:2013}
B.~Ayg{\"u}n and A.~Soysal, ``Capacity bounds on mimo relay channel with
  covariance feedback at the transmitters,'' \emph{IEEE Trans. Veh. Technol.},
  vol.~62, no.~5, pp. 2042--2051, May 2013.

\bibitem{moh4}
M.~Samavat, F.~Hosseinigoki, and S.~Talebi, ``Performance improvement of
  mimo-ofdm block codes by achieving a suboptimum permutation distance,''
  \emph{Majlesi Journal of Telecommunication Devices}, vol.~2, no.~4, 2013.

\bibitem{moh3}
M.~Samavat, G.~F. Hosseini, and S.~Talebi, ``Alamouti coding scheme for
  cooperative relay networks with full duplex relaying,'' in \emph{Iran
  Workshop on Communication and Information Theory, {IWCIT} 2013, Tehran, Iran,
  May 8-9, 2013}, 2013, pp. 1--4.

\bibitem{Khisti:TIT:2010}
A.~Khisti and G.~W. Wornell, ``Secure transmission with multiple antennas i:
  the misome wiretap channel,'' \emph{IEEE Trans. Inf. Theory}, vol.~56, no.~7,
  pp. 3088--3104, Jul. 2010.

\bibitem{Huang:TSP:2011}
J.~Huang and A.~L. Swindlehurst, ``Cooperative jamming for secure
  communications in mimo relay networks,'' \emph{IEEE Trans. Signal Process},
  vol.~59, no.~10, pp. 4871--4884, Oct. 2011.

\bibitem{Hassibi:TIT:2011}
F.~E. Oggier and B.~Hassibi, ``The secrecy capacity of the mimo wiretap
  channel,'' \emph{IEEE Trans. Inf. Theory}, vol.~57, no.~8, pp. 4961--4972,
  Aug. 2011.

\bibitem{Vahidian:WCL:2015}
S.~Vahidian, S.~A{\"{\i}}ssa, and S.~Hatamnia, ``Relay selection for
  security-constrained cooperative communication in the presence of
  eavesdropper's overhearing and interference,'' \emph{{IEEE} Wireless Commun.
  Lett.}, vol.~4, no.~6, pp. 577--580, Dec. 2015.

\bibitem{Petropulu:TSP:2010}
L.~Dong, Z.~Han, A.~P. Petropulu, and H.~V. Poor, ``Improving wireless physical
  layer security via cooperating relays,'' \emph{IEEE Trans. Signal Process.},
  vol.~58, no.~3, pp. 1875--1888, Mar. 2010.

\bibitem{Shen:Conf:2013}
X.~W. Y.~Zou and W.~Shen, ``Intercept probability analysis of cooperative
  wireless networks withbest relay selection in the presence of eavesdropping
  attack,'' \emph{inProc. IEEE Intern. Conf. Commun. (ICC)}, pp. 1--5,
  Budapest, Hungary, Jun. 2013.

\bibitem{moh2}
M.~Samavat, A.~Morsali, and S.~Talebi, ``Delay-interleaved cooperative relay
  networks,'' \emph{{IEEE} Communications Letters}, vol.~18, no.~12, pp.
  2137--2140, 2014.

\bibitem{Erkip:Trans:2011}
M.~Yuksel, X.~Liu, and E.~Erkip, ``A secure communication game with a relay
  helping the eavesdropper,'' \emph{IEEE Trans. Inf. Forensics and Security},
  vol.~6, no. 3-1, pp. 818--830, 2011.

\bibitem{Oohama:2006}
Y.~Oohama, ``Relay channels with confidential messages,'' \emph{CoRR}, vol.
  abs/cs/0611125, 2006.

\bibitem{Fawaz:TWC:2015}
F.~S. Al{-}Qahtani, C.~Zhong, and H.~M. Alnuweiri, ``Opportunistic relay
  selection for secrecy enhancement in cooperative networks,'' \emph{{IEEE}
  Trans. Commun.}, vol.~63, no.~5, pp. 1756--1770, May 2015.

\bibitem{moh1}
A.~Sureshbabu, M.~Samavat, X.~Li, and C.~Tepedelenlioglu, ``Outage probability
  of multi-hop networks with amplify-and-forward full-duplex relaying,''
  \emph{{IET} Communications}, vol.~12, no.~13, pp. 1550--1554, 2018.

\bibitem{ElGamal:TIF:2008}
L.~Lai and H.~E. Gamal, ``The relay-eavesdropper channel: Cooperation for
  secrecy,'' \emph{IEEE Trans. Inf. Theory}, vol.~54, no.~9, pp. 4005--4019,
  Sep. 2008.

\bibitem{Jeong:TSP:2011}
C.~Jeong and I.-M. Kim, ``Optimal power allocation for secure multicarrier
  relay systems,'' \emph{IEEE Trans. Signal Process.}, vol.~59, no.~11, p.
  5428–5442, Nov. 2011.

\bibitem{Thompson:TWC:2009}
I.~Krikidis, J.~S. Thompson, and S.~McLaughlin, ``Relay selection for secure
  cooperative networks with jamming,'' \emph{IEEE Trans. Wireless Commun.},
  vol.~8, no.~10, pp. 5003--5011, Oct. 2009.

\bibitem{Liang:SPL:2015}
H.~Hui, A.~L. Swindlehurst, G.~Li, , and J.~Liang, ``Secure relay and jammer
  selection for physical layer security,'' \emph{IEEE Signal Process. Lett.},
  vol.~22, no.~8, pp. 1147--1151, 2015.

\bibitem{Chen:TIF:2012}
L.~S. Z. H. a. B.~J. J.~Chen, R.~Zhang, ``Joint relay and jammer selection for
  secure two-way relay networks,'' \emph{IEEE Trans. Inf. Forens. Security},
  vol.~7, no.~1, pp. 310--320, Feb. 2012.

\bibitem{IESe}
H.~Liu, P.~L. Yeoh, K.~J. Kim, P.~V. Orlik, and H.~V. Poor, ``Secrecy
  performance of finite-sized in-band selective relaying systems with
  unreliable backhaul and cooperative eavesdroppers,'' \emph{IEEE Journal on
  Sel. Areas in Commun.}, pp. 245--256, Mar 2018.

\bibitem{qaa}
Y.~Cai and C.~Zhang, ``Physical layer security in wireless-powered networks
  with untrusted relays,'' in \emph{2017 IEEE 17th International Conf. on
  Commun. Technol.(ICCT)}, Oct 2017, pp. 771--776.

\bibitem{qaaa}
X.~Ding, T.~Song, Y.~Zou, X.~Chen, and L.~Hanzo, ``Security-reliability
  tradeoff analysis of artificial noise aided two-way opportunistic relay
  selection,'' \emph{IEEE Trans. on Veh. Technol.}, vol.~66, no.~5, pp.
  3930--3941, May 2017.

\bibitem{qaaaa}
H.~Fang, L.~Xu, and X.~Wang, ``Coordinated multiple-relays based physical-layer
  security improvement: A single-leader multiple-followers stackelberg game
  scheme,'' \emph{IEEE Trans. on Inf. Forensics and Security}, vol.~13, no.~1,
  pp. 197--209, Jan 2018.

\bibitem{qbb}
D.~Tubail, M.~El-Absi, S.~S. Ikki, W.~Mesbah, and T.~Kaiser, ``Artificial
  noise-based physical-layer security in interference alignment multipair
  two-way relaying networks,'' \emph{IEEE Access}, vol.~6, pp.
  19\,073--19\,085, 2018.

\bibitem{qbbb}
L.~Wu, L.~Yang, J.~Chen, and M.~S. Alouini, ``Physical layer security for
  cooperative relaying over generalized-k fading channels,'' \emph{IEEE
  Wireless Commun. Lett.}, pp. 234--237, 2018.

\bibitem{qbbbb}
D.~Chen, Y.~Cheng, W.~Yang, J.~Hu, and Y.~Cai, ``Physical layer security in
  cognitive untrusted relay networks,'' \emph{IEEE Access}, vol.~6, pp.
  7055--7065, 2018.

\bibitem{Larsson:TWC:2009}
M.~N. Khormuji and E.~G. Larsson, ``Cooperative transmission based on
  decode-and-forward relaying with partial repetition coding,'' \emph{{IEEE}
  Trans. Wireless Commun.}, vol.~8, no.~4, pp. 1716--1725, Apr. 2009.

\bibitem{Zurita:SPL:2012}
N.~Romero{-}Zurita, M.~Ghogho, and D.~C. McLernon, ``Outage probability based
  power distribution between data and artificial noise for physical layer
  security,'' \emph{{IEEE} Signal Process. Lett.}, vol.~19, no.~2, pp. 71--74,
  Feb. 2012.

\bibitem{McKay:TVT:2013}
X.~Zhang, X.~Zhou, and M.~R. McKay, ``On the design of artificial-noise-aided
  secure multi-antenna transmission in slow fading channels,'' \emph{{IEEE}
  Trans. Veh. Technol.}, vol.~62, no.~5, pp. 2170--2181, June 2013.

\bibitem{6951465}
Q.~Li, Y.~Yang, W.~K. Ma, M.~Lin, J.~Ge, and J.~Lin, ``Robust cooperative
  beamforming and artificial noise design for physical-layer secrecy in af
  multi-antenna multi-relay networks,'' \emph{IEEE Trans. Signal Process.},
  vol.~63, no.~1, pp. 206--220, Jan 2015.

\bibitem{7752562}
W.~Liu, M.~Li, G.~Ti, X.~Tian, and Q.~Liu, ``Transmit filter and artificial
  noise aided physical layer security for ofdm systems,'' in \emph{Proc. 8th
  International Conf. Wireless Commun. Signal Process. (WCSP)}, Oct. 2016, pp.
  1--5.

\bibitem{Wang17}
W.~Wang, K.~C. Teh, and K.~H. Li, ``Artificial noise aided physical layer
  security in multi-antenna small-cell networks,'' \emph{IEEE Trans. Inf.
  Forens. Security}, vol.~12, no.~6, pp. 1470--1482, June 2017.

\bibitem{7417842}
L.~Zhang, H.~Zhang, D.~Wu, and D.~Yuan, ``Improving physical layer security for
  miso systems via using artificial noise,'' in \emph{Proc. IEEE Global Commun.
  Conf. (GLOBECOM)}, Dec 2015, pp. 1--6.

\bibitem{Shen:TVT:2014}
Y.~Zou, X.~Wang, W.~Shen, and L.~Hanzo, ``Security versus reliability analysis
  of opportunistic relaying,'' \emph{IEEE Trans. Veh. Technol.}, vol.~63,
  no.~6, pp. 2653--2661, June 2014.

\bibitem{B.Champagne:TCOM:2015}
Y.~Zou, B.~Champagne, W.-P. Zhu, and L.~Hanzo, ``Relay-selection improves the
  security-reliability trade-off in cognitive radio systems,'' \emph{IEEE
  Trans. Commun.}, vol.~63, no.~1, pp. 215--228, Jan. 2015.

\bibitem{Bao:TWC:2013}
V.~N.~Q. Bao, N.~Linh-Trung, and M.~Debbah, ``Relay selection schemes for
  dual-hop networks under security constraints with multiple eavesdroppers,''
  \emph{IEEE Trans. Wireless Commun.}, vol.~12, no.~12, pp. 6076--6085, Nov.
  2013.

\bibitem{7996473}
S.~Vahidian, M.~Najafi, M.~Najafi, and F.~S. Al-Qahtani, ``Power allocation and
  cooperative diversity in two-way non-regenerative cognitive radio networks,''
  in \emph{2017 IEEE International Conference on Communications (ICC)}, May
  2017, pp. 1--7.

\bibitem{hatamnia2017network}
S.~Hatamnia, S.~Vahidian, S.~A{\"\i}ssa, B.~Champagne, and M.~Ahmadian-Attari,
  ``Network-coded two-way relaying in spectrum-sharing systems with
  quality-of-service requirements,'' \emph{IEEE Transactions on Vehicular
  Technology}, vol.~66, no.~2, pp. 1299--1312, Feb. 2017.

\bibitem{Vahidian:2015:WPC}
M.~Najafi, M.~Ardebilipour, E.~Soleimani{-}Nasab, and S.~Vahidian, ``Multi-hop
  cooperative communication technique for cognitive {DF} and {AF} relay
  networks,'' \emph{Wireless Personal Communications}, vol.~83, no.~4, pp.
  3209--3221, 2015.

\bibitem{Negi:TWC:2008}
S.~Goel and R.~Negi, ``Guaranteeing secrecy using artificial noise,''
  \emph{IEEE Trans. Wireless Commun.}, vol.~7, no.~6, pp. 2180--2189, June
  2008.

\bibitem{Shannon19488}
C.~E. Shannon, ``A mathmatical theory of communication,'' \emph{Bell System
  Technical Journal}, vol.~27, no.~6, pp. 379--423, 1948.

\bibitem{Bletsas2006}
A.~Bletsas, A.~Khisti, D.~Reed, and A.~Lippman, ``A simple cooperative
  diversity method based on network path selection,'' \emph{{IEEE} J. Sel.
  Areas Commun.}, vol.~24, no.~3, pp. 659-- 672, March 2006.

\bibitem{Zou.Trans.2010}
Y.~Zou, J.~Zhu, B.~Zheng, and Y.-D. Yao, ``An adaptive cooperation diversity
  scheme with best-relay selection in cognitive radio networks,'' \emph{IEEE
  Trans. Signal Process.}, vol.~58, no.~10, pp. 5438--5445, Oct. 2010.

\bibitem{saeedtvt2017}
S.~Vahidian, E.~Soleimani-Nasab, S.~A{\"\i}ssa, and M.~Ahmadian-Attari,
  ``Bidirectional af relaying with underlay spectrum sharing in cognitive radio
  networks,'' \emph{IEEE Transactions on Vehicular Technology}, vol.~66, no.~3,
  pp. 2367--2381, Mar. 2017.

\bibitem{Bletsas:TWC:2007}
H.~S. A.~Bletsas and M.~Z. Win, ``Cooperative communications with
  outage-optimal opportunistic relaying,'' \emph{IEEE Trans. Wireless Commun.},
  vol.~6, no.~9, p. 3450–3460, Jun. 2007.

\bibitem{}


\bibitem{Ikhlef:GLOBE:2011}
A.~Ikhlef, D.~S. Michalopoulos, and R.~Schober, ``Buffers improve the
  performance of relay selection,'' in \emph{Proc. of the Global Commun. Conf.,
  {GLOBECOM} , Dec. , Houston, Texas, {USA}}, 2011, pp. 1--6.

\bibitem{Bloch:TIF:2008}
M.~R. Bloch, J.~Barros, M.~R.~D. Rodrigues, and S.~W. McLaughlin, ``Wireless
  information-theoretic security,'' \emph{IEEE Trans. Inf. Theory}, vol.~54,
  no.~6, pp. 2515--2534, June 2008.

\bibitem{Papoulis:1984}
A.~Papoulis, \emph{Probability, Random variables, and Stochastic
  Processes.}\hskip 1em plus 0.5em minus 0.4em\relax New York: McGraw-Hill,
  1991.

\bibitem{Wolfram:website}
Wolfram, ``The {W}olfram functions site,'' \emph{Available:
  http://functions.wolfram.com, 2012.}

\end{thebibliography}

\end{document}